\documentclass[%
 reprint,
 superscriptaddress,
 amsmath,amssymb,
 prl,
]{revtex4-2}

\usepackage{graphicx}
\usepackage{dcolumn}
\usepackage{bm}
\usepackage{gensymb}
\usepackage{svg}
\usepackage{afterpage}
\usepackage{physics}
\usepackage{tabularx}
\usepackage{xr}
\usepackage{natbib}
\usepackage{amsmath}
\usepackage{amsfonts}
\usepackage[T1]{fontenc}
\usepackage[utf8]{inputenc}

\usepackage{xr}
\begin{document}

\preprint{APS/123-QED}

\title{Spatial self-organization driven by temporal noise}

\author{Satyam Anand}
\email{sa7483@nyu.edu}
\affiliation{Courant Institute of Mathematical Sciences, New York University, New York, NY 10003, USA}
\affiliation{\mbox{Center for Soft Matter Research, Department of Physics, New York University, New York, NY 10003, USA}}

\author{Guanming Zhang}
\affiliation{\mbox{Center for Soft Matter Research, Department of Physics, New York University, New York, NY 10003, USA}}
\affiliation{\mbox{Simons Center for Computational Physical Chemistry, Department of Chemistry, New York University, New York, NY 10003, USA}}

\author{Stefano Martiniani}
\email{sm7683@nyu.edu}
\affiliation{Courant Institute of Mathematical Sciences, New York University, New York, NY 10003, USA}
\affiliation{\mbox{Center for Soft Matter Research, Department of Physics, New York University, New York, NY 10003, USA}}
\affiliation{\mbox{Simons Center for Computational Physical Chemistry, Department of Chemistry, New York University, New York, NY 10003, USA}}
\affiliation{\mbox{Center for Neural Science, New York University, New York, NY 10003, USA}}

\begin{abstract}

The counterintuitive emergence of order from noise is a central phenomenon in science, ranging from pattern formation and synchronization to order-by-disorder in frustrated systems. 
While large-scale spatial self-organization induced by local spatial noise is well studied, whether temporal noise can also drive such organization remains an open question. 
Here, by studying interacting particle systems, we show that temporally correlated noise can lead to a self-organized state with suppressed long-range density fluctuations, or hyperuniformity.
Further, we develop a fluctuating hydrodynamic theory that quantitatively explains the origin of this phenomenon.
Finally, by casting the dynamics as a stochastic optimization problem, we show that temporal correlations lead to better solutions, akin to perturbed gradient descent in neural networks---where noise is injected during training to escape poor minima. This reveals a striking correspondence between perturbed gradient descent dynamics on the energy landscapes of particle systems and the loss landscapes of neural networks.
Our study establishes temporal correlations as a novel mechanism for noise-driven self-organization, with broad implications for self-assembling materials, biological systems, and optimization algorithms that leverage temporal noise for applications like differentially private learning.

\end{abstract}

\maketitle

Noise-driven spatiotemporal order is a well-known phenomenon in science and mathematics \cite{sagues2007spatiotemporal, garcia2012noise}, with examples ranging from Rayleigh–Bénard convection \cite{rayleigh1916lix, benard1900tourbillons} and the Belousov–Zhabotinsky reaction \cite{kuhnert1986new} to swarming \cite{jhawar2020noise}, and order-by-disorder in frustrated magnets \cite{villain1980order}. Such phenomena can be broadly divided into three classes depending on the driving mechanism---spatial, temporal, or spatiotemporal noise \cite{sagues2007spatiotemporal}. Although spatial noise driven self-organization in interacting many-body systems is well studied \cite{corte2008random, hexner2017enhanced, hexner2017noise, wilken2021random, wilken2023dynamical, tjhung2015hyperuniform, lei2023random, zhang2024absorbing, anand2025emergent}, it remains unclear whether, and how, purely temporal noise can generate large-scale spatial structures.

In thermal equilibrium, fluctuations are typically white ($\delta$-correlated) and obey the fluctuation-dissipation theorem (FDT). Temporally correlated noise that violates the FDT is an intrinsically non-equilibrium phenomenon, appearing naturally across a wide variety of systems, including colloids in active baths \cite{wu2000particle, dabelow2019irreversibility, maggi2014generalized}, polymers subject to active forces \cite{goychuk2024delayed, brahmachari2024temporally}, active harmonic chains \cite{ikeda2024harmonic, marconi2024active}, surface growth processes \cite{katzav2004kardar}, and anomalous diffusion \cite{ikeda2024harmonic, eliazar2009unified}. Unlike one-body systems driven by temporal noise that have been well studied, interacting many-body systems with temporal noise remain underexplored, with some exceptions \cite{ikeda2024harmonic}. We thus ask: Can temporal noise with short-memory (short-range temporal correlations) \textit{alone} drive large-scale spatial self-organization?

This question finds a surprising parallel in optimization and machine learning. Optimization algorithms used to train neural networks, such as perturbed gradient descent (PGD) \cite{song2013stochastic, orvieto2022anticorrelated, orvieto2023explicit, choquette2023correlated, jin2021nonconvex}, deliberately inject temporal noise during training, for applications such as differential privacy, which ensures that individual training data cannot be identified from the trained model \cite{ponomareva2023dp, dwork2006calibrating}. 
It is empirically known that temporal correlations in the noise help such optimizers navigate to ``good'' minima, i.e., ones that yield improved generalization on unseen data \cite{orvieto2023explicit, choquette2023correlated}.
The success of temporal noise-driven navigation of neural network loss landscapes poses an intriguing question from a physics perspective: Does PGD dynamics on energy landscapes of particle systems also discover better minima?

Here, combining simulations and theory, we address these questions by studying interacting particle systems driven by short-memory temporally correlated noise. Our contributions are threefold. 
First, we show that temporal anti-correlations lead to large-scale spatial self-organization, in particular, the suppression of long-wavelength density fluctuations below a crossover length scale, which diverges when the noise becomes strongly anti-correlated, leading to hyperuniformity \cite{torquato2018hyperuniform}.
Second, we develop a fluctuating hydrodynamic theory by directly coarse-graining the microscopic dynamics, that quantitatively explains the emergent self-organization and crossover behavior.
Third, by recasting the dynamics as an optimization problem, we demonstrate that temporal noise biases the dynamics toward ``good'' (deeper and flatter) minima, thus uncovering a striking correspondence between PGD dynamics in two disparate systems---neural networks and particle systems.
Our study offers a new paradigm for self-assembling hyperuniform structures, provides a physical basis for designing novel differentially private learning algorithms, and sheds light on the role of temporal fluctuations, ubiquitous in diverse fields such as chromatin folding \cite{goychuk2024delayed, brahmachari2024temporally}, active matter \cite{wu2000particle, dabelow2019irreversibility, maggi2014generalized}, ecology \cite{zanchetta2025emergence, crocker2025microbial}, and neuroscience \cite{xu2019effects, jedynak2017temporally}.

\begin{figure*}[htpb!]
    \centering
    \includegraphics[width=\linewidth]{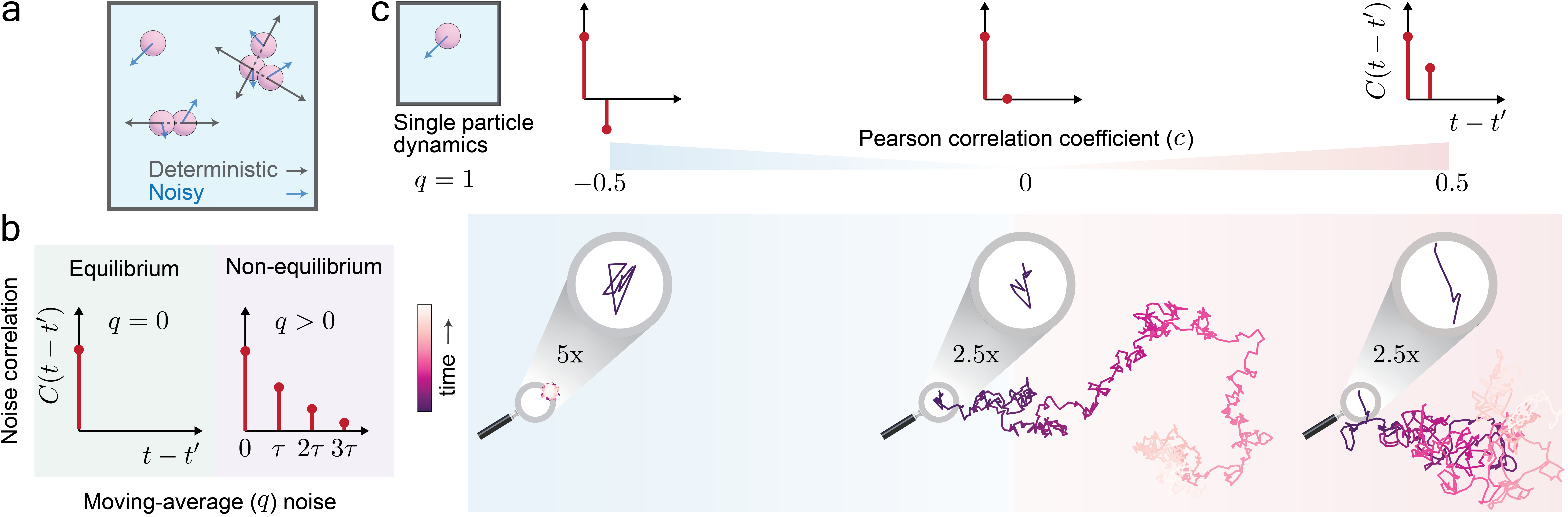}
    \caption{\small \textbf{Interacting particle systems driven by temporally correlated noise.} (a) Schematic of the system showing passive particles immersed in a non-equilibrium bath (light blue). Black arrows denote deterministic pairwise interactions, while blue arrows represent noise from the bath. (b) Schematic of the correlation structure of the moving-average $q$ (MA($q$)) noise, where $q$ denotes the number of time steps with non-zero correlations. For $q=0$, the noise is white, reducing the system to an equilibrium interacting Brownian particle system. For $q>0$, the noise is temporally correlated, driving the system out of equilibrium. (c) Dynamics of a single particle driven by MA($1$) noise. Noise correlation structures for strongly anti-correlated ($c=-1/2$), uncorrelated ($c=0$), and strongly correlated ($c=1/2$) cases (Top). Exemplar particle trajectories from simulations for each case (Bottom). Here, $c$ is the Pearson correlation coefficient between noise at times $t$ and $t+1$ (Eq.~\ref{eq:pcc_maq}). Zoomed in sections of the particle trajectories highlight that strongly correlated noise tends to produce persistent motion in one direction, whereas strongly anti-correlated noise causes frequent reversals. 
}
    \label{fig:figure1}
\end{figure*}

\subsection{Setup}

Our system consists of $N$ interacting spherical particles of radius $R$ in $d$ spatial dimensions, following discrete-time dynamics. The position of particle $i$ at time-step $m$ evolves as,
\begin{equation}
  \mathbf{x}_{i}^{m} = \mathbf{x}_{i}^{m-1} - \alpha \sum_{j \in \Gamma_i^{m-1}} \nabla_{i} V_{ji}^{m-1} + \boldsymbol{\zeta}_i^{m},
\label{eq:discrete_pgd}
\end{equation}
where $\nabla_{i} = \nabla_{\mathbf{x}_{i}}$, $\alpha$ is the learning rate with units of $\text{length}/\text{force}$, $V_{ji}^{m-1} = V(|\mathbf{x}^{m-1}_{j} - \mathbf{x}^{m-1}_{i}|)$ is a pairwise interaction potential between particles $i$ and $j$, and $\Gamma_i^{m-1} = \{j \mid |\mathbf{x}_j^{m-1} - \mathbf{x}_i^{m-1}| < 2R,\ j \neq i \}$ is the set containing all particles that overlap with particle $i$ at time-step $m-1$ (Fig.~\ref{fig:figure1}a). 
While $V_{ji}$ can be any potential (short- or long-range), here, we consider short-range, repulsive potentials (Eq.~\ref{eq:v_ji_family}, Methods).

The noise term in Eq.~\ref{eq:discrete_pgd} is given by,
\begin{equation}
 \boldsymbol{\zeta}_i^{m} = \frac{\sigma}{\Theta} \sum_{k=0}^{q} \theta_k \boldsymbol{\xi}_i^{m-k},
\label{eq:pgd_noise}
\end{equation}
where $\Theta = \sqrt{\sum_{k=0}^{q} \theta_k^2}$, $\boldsymbol{\xi}_i^{m}$ is a standard Gaussian, white noise with $\left< {\xi}_{i, \alpha}^{m} \right> = 0$, and $\left< {\xi}_{i, \alpha}^{m} {\xi}_{j, \beta}^{n} \right> = \delta_{ij} \delta_{\alpha \beta} \delta^{mn}$. $\sigma$ has units of $\text{length}$ and controls the noise magnitude. $\theta_0 = 1$, and $\{\theta_k\}_{k=1}^{q} \in [-1, 1]$ are scalar parameters that weight noise from the previous $q$ time steps. The Pearson correlation coefficient between ${\zeta}_{i, \alpha}^{m}$ and ${\zeta}_{i, \alpha}^{m-h}$ is given by,
\begin{equation}
  c_h =  \begin{cases}
  \frac{1}{\Theta^2} \sum_{k=0}^{q-h} \theta_k \theta_{k+h}, & \text{if}\ 1 \leq h \leq q, \\
  0, & \text{if}\ h > q.  
  \end{cases}
 \label{eq:pcc_maq}
\end{equation} 

\begin{figure*}[htpb!]
    \centering
    \includegraphics[width=\linewidth]{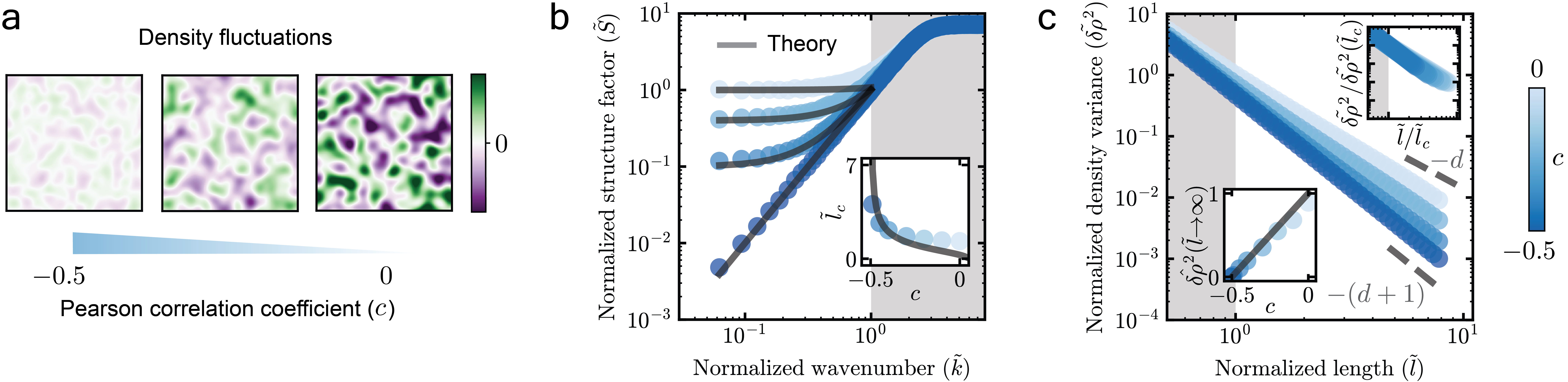}
    \caption{\small \textbf{Emergent long-range structure.} (a) Coarse-grained density fluctuations $\delta \rho(c)/|\delta \rho_{\text{avg}}(c=0)|$ from particle simulations, where $c$ is the noise correlation coefficient and $\delta \rho_{\text{avg}}$ is the average density fluctuation of the whole system. Panels show $c=-0.5$, $-0.4$, and $0$ (left to right). The system was coarse-grained with a Gaussian kernel of width $80R$, where $R$ is the particle radius. As $c$ decreases from $0$ (uncorrelated) to $-0.5$ (strongly anti-correlated), density fluctuations are progressively suppressed. (b) Normalized radially averaged structure factor $\Tilde{S}(\Tilde{k})$ versus normalized wave number $\Tilde{k}$. Here, $\tilde{S} =  S(k)/S_{0}(2\pi/L)$, where $S_{0}(2\pi/L)$ is the structure factor for $c=0$ at $k = 2\pi/L$, and $L$ is the simulation box side length. $\tilde{k} = k/k_0$, where $k_0$ is the value at which $\tilde{S}(k_0) = 1$ for $c=-0.5$. Solid black lines show predictions of Eq.~\ref{eq:struc_fact} for different $c$. Inset: Normalized crossover length scale ($l_c/l_0 = \tilde{l}_c = 1/\tilde{k}_c$) versus $c$. The normalized crossover wavenumber ($\tilde{k}_c$) is obtained in simulations from the intersection, on a log-log plot, between a slope $0$ fit near $\tilde{k} \to 0$, and a slope $2$ fit near $\tilde{k} \approx 1$. Solid black line in the inset shows the prediction of Eq.~\ref{eq:l_c}. Gray shaded regions indicate short-range behavior ($\tilde{k}>1$). (c) Normalized number density $\tilde{\delta \rho}^2 (\tilde{l})$ versus normalized hypersphere diameter ($\Tilde{l}$) used for measuring density fluctuations. $\tilde{\delta \rho}^2 (\tilde{l}) = \delta \rho^2 (l)/\delta \rho^2 (l_0)$ where $\delta \rho^2 (l_0)$ is the density variance for $c=0$ at $l=l_0$, and $\tilde{l}=l/l_0$, with $l_0 = 2\pi/k_0$. Top inset: Data collapse of density variances upon rescaling $\tilde{l}$ by $\tilde{l}_c$ and $\tilde{\delta \rho}^2 (\tilde{l})$ by $\tilde{\delta \rho}^2 (\tilde{l}_c)$. Bottom inset: Infinite wavelength density fluctuations $\hat{\delta \rho}^2 (\tilde{l} \to \infty)$ versus $c$, where $\hat{\delta \rho}^2 (c) = [\tilde{\delta \rho}^2 (c) - \tilde{\delta \rho}^2 (c=-0.5)]/\tilde{\delta \rho}^2 (c=0)$. Solid black line denotes the prediction $1+2c$ from Eq.~\ref{eq:struc_fact} in the $\tilde{k} \to 0$ limit. Gray shaded regions indicate short-range behavior ($\tilde{l}<1$). Circles in (b) and (c) denote particle simulations.
}
    \label{fig:figure2}
\end{figure*}

The noise given by Eq.~\ref{eq:pgd_noise} is called as the moving-average (MA($q$)) noise, where $q$ is the number of time-steps up to which the noise has non-zero correlations (Eq.~\ref{eq:pcc_maq}, Fig.~\ref{fig:figure1}b). MA($q$) noise is widely studied in time-series analysis and forecasting in diverse fields such as climate, economics, and finance \cite{shumway2006time}, and has several advantages: (i) since correlations are zero after time $q$, it can model noise with tunable memory by adjusting $q$, (ii) it can approximate arbitrary correlation structure by tuning parameters $\{\theta_k\}_{k=1}^{q}$.

Where could MA($q$)-type noise typically arise in physical systems? 
It can occur in active systems where an initial random force, or ``kick'', generated by an active (energy-consuming) process, is followed by an ``echo'' after a time delay \cite{goychuk2024delayed}. 
For instance, a polymer segment may receive an active push from a molecular motor, after which its elasticity may cause it to snap back. This recoil acts as a negative echo, producing a push-pull sequence localized at a single site. 
Another example could be chemical reaction cycles in which a protein binds to a monomer (the initial kick), undergoes a conformational change while bound, and then unbinds after a delay \cite{goychuk2024delayed}.

For simplicity, we consider MA($1$) noise ($q=1$) in the main text (Fig.~\ref{fig:figure1}c); however, all our numerical and analytical results hold true for the general MA($q$) noise (see Supplementary Information (SI)). The MA($1$) noise is characterized solely by the lag-$1$ noise correlation coefficient $c_1 \in [-1/2,1/2]$, which we denote by $c$ hereafter (Eq.~\ref{eq:pcc_maq}, see SI Eq.~\ref{eq:sum_c_bounds} for a discussion of the bounds). Fig.~\ref{fig:figure1}c shows representative trajectories of a single particle driven by MA($1$) noise, highlighting that strongly correlated noise tends to produce persistent motion in one direction, whereas strongly anti-correlated noise causes frequent reversals.
We report all our results when the system reaches a steady-state, i.e., when the total energy $E = \sum_i \sum_{j \geq i} V_{ij}$ fluctuates around a time-independent constant value.

\subsection{Emergent long-range structure}

We quantify the long-range structure in particle simulations by measuring the radially averaged structure factor $S(k)$ and the number density variance $\delta \rho^2 (l)$, where $k = 2 \pi /l$ is the wave number, and $l$ is the diameter of the sphere used to measure density fluctuations (Methods). Since we are interested in the long-range behavior, we study all properties above a length scale $l_0$ (see Fig.~\ref{fig:figure2} caption for definition). Moving forward, we normalize all quantities and get $\tilde{l}$, $\tilde{k}$, $\tilde{S}(\tilde{k})$, and $\tilde {\delta \rho}^2 (\tilde{l})$, see Fig.~\ref{fig:figure2} caption. 

As the noise becomes more anti-correlated, large-scale density fluctuations are suppressed (Fig.~\ref{fig:figure2}a). The system self-organizes to suppress density fluctuations below a normalized crossover length scale $\tilde{l}_c$. For $\tilde{l} < \tilde{l}_c$, the structure factor follows a power law behavior ($\tilde{S}(\tilde{k}) \sim k^2$), while for $\tilde{l} > \tilde{l}_c$ the structure factor is constant ($\tilde{S}(\tilde{k}) \sim \text{const.}$) (Fig.~\ref{fig:figure2}b). Similarly, the density fluctuations decay as $\tilde{\delta \rho} ^2 \sim \tilde{l}^{-(d+1)}$ for $\tilde{l} < \tilde{l}_c$, whereas for $\tilde{l} > \tilde{l}_c$, density fluctuations follow $\tilde{\delta \rho} ^2 \sim \tilde{l}^{-d}$ (Fig.~\ref{fig:figure2}c). The crossover length scale $l_c$ increases monotonically as $c$ decreases, and ultimately diverges when $c = -1/2$, meaning that the density fluctuations are suppressed to larger length scales as the noise becomes more anti-correlated (Fig.~\ref{fig:figure2}b inset). The infinite wavelength density fluctuations $\tilde{\delta \rho}^2 (\tilde{l} \to \infty) \propto \tilde{S}(\tilde{k} \to 0)$ also decrease as $c$ decreases (Fig.~\ref{fig:figure2}c inset). Finally, when the noise is maximally anti-correlated ($c=-1/2$), the crossover length $\tilde{l_c} \to \infty$, the density fluctuations decay at the fastest possible rate, $\tilde{\delta \rho} ^2 (\tilde{l} \to \infty) \sim \tilde{l}^{-(d+1)}$, and $\tilde{S}(\tilde{k} \to 0) \sim \tilde{k}^{2}$, corresponding to strong (class I \cite{torquato2018hyperuniform}) hyperuniformity (Fig.~\ref{fig:figure2}b, c).

The long-range behavior depends primarily on the noise correlation $c$, and is qualitatively independent of all other parameters such as the learning rate $\alpha$, pairwise potential $V_{ij}$, noise magnitude $\sigma$, volume fraction $\phi$, noise correlation cutoff time $q$, and remarkably, the spatial dimension $d$ (SI Fig.~\ref{fig:figure1_si}).

\subsection{Fluctuating hydrodynamics}

We now ask: What is the mechanism behind this temporal noise-driven spatial self-organization?
To answer this, we first approximate the discrete-time dynamics (Eq.~\ref{eq:discrete_pgd}) by a continuous-time stochastic differential equation (SDE). To do so, we replace the MA($1$) noise by a Gaussian, colored noise having identical power spectral density to get an overdamped Langevin equation for the evolution of the position of particle $i$ (see SI Sec.~I for derivation),
\begin{equation}
 \frac{d\mathbf{x}_{i}(t)}{dt}
 =  - \frac{1}{\gamma} \sum_{j \in \Gamma_i^t} \nabla_{i} V_{ji}
 + \sqrt{2D} \ \boldsymbol{\eta}_{i}(t),
 \label{eq:cont_pgd} 
\end{equation}
where $\gamma = \tau/\alpha$ is the friction coefficient, $D = \sigma^2/2 \tau$ is the diffusion constant, and $\tau$ is the time scale corresponding to a discrete time-step. $\boldsymbol{\eta}_{i}(t)$ is a Gaussian colored noise with mean $\left< {\eta}_{i, \alpha}(t) \right> = 0$, and covariance matrix given by,
\begin{align}
\left< {\eta}_{i,\alpha} (t) \, {\eta}_{j,\beta} (t') \right> &= \delta_{ij} \delta_{\alpha \beta} \left[ \delta(t-t') + c \, \delta(|t-t'| - \tau) \right].
\label{eq:cov_eta}
\end{align}

We coarse-grain the continuous-time SDE to formulate a theory for the evolution of the density field $\rho (\mathbf{x}, t)$. We use Dean's method---originally formulated for systems with additive, white noise \cite{dean1996langevin}, and later extended to systems with multiplicative \cite{bertin2013mesoscopic, solon2015active, donev2014dynamic, anand2025emergent}, and colored \cite{russo2020memory, feng2021effective} noise---to get (see SI Sec.~III for derivation),
\begin{align}
    \frac{\partial \rho(\mathbf{x}, t)}{\partial t} 
    &= \underbrace{\frac{1}{\gamma} \nabla \cdot \left[ \rho(\mathbf{x}, t)  \int \rho(\mathbf{y}, t) \nabla V(\mathbf{x},\mathbf{y}) d\mathbf{y}   \right]}_\text{drift term} \nonumber \\  
    &\quad + \underbrace{D \nabla^2  \rho(\mathbf{x}, t)}_\text{diffusion term} \nonumber \\  
    &\quad + \underbrace{\sqrt{2D} \ \nabla \cdot \left[ \sqrt{\rho(\mathbf{x}, t)} \, \boldsymbol{\psi}(\mathbf{x}, t) \right]}_\text{noise term},  
    \label{eq:generic_drho}
\end{align}
where $\nabla = \nabla_{\mathbf{x}}$, and $V(\mathbf{x}, \mathbf{y})$ is the continuous form of $V_{ji}$, given by substituting $\mathbf{x}_i$ and $\mathbf{x}_j$ by $\mathbf{x}$ and $\mathbf{y}$ in Eq.~\ref{eq:v_ji_family}.

The drift term in Eq.~\ref{eq:generic_drho} is of the form $-\nabla \cdot (\rho \mathbf{v})$, where the velocity $\mathbf{v}(\mathbf{x})$ originates from the deterministic (first) term in Eq.~\ref{eq:cont_pgd}, and represents the mean interaction force $- \langle \nabla V(\mathbf{x}, \mathbf{y}) \rangle_{\rho(\mathbf{y})}$ exerted at location $\mathbf{x}$, scaled by the friction coefficient $\gamma$.

The noise term in Eq.~\ref{eq:generic_drho} originates from the noise (second) term in Eq.~\ref{eq:cont_pgd}, where $\boldsymbol{\psi}(\mathbf{x},t)$ is a Gaussian noise field with mean $\left< {\psi}_{\alpha}(\mathbf{x}, t) \right> = 0$, and covariance matrix $\left< {\psi}_{\alpha}(\mathbf{x}, t) \, {\psi}_{\beta}(\mathbf{y}, t')\right> = \delta_{\alpha \beta} \delta(\mathbf{x} - \mathbf{y}) \left[ \delta(t-t') + c \, \delta(|t-t'| - \tau) \right]$. $c$ is the Pearson correlation coefficient between ${\psi}_{\alpha}(\mathbf{x}, t)$ and ${\psi}_{\alpha}(\mathbf{x}, t + \tau)$. 

Notice that setting $c=0$ in Eq.~\ref{eq:cov_eta} reduces the system (Eq.~\ref{eq:cont_pgd}) to an equilibrium system of interacting Brownian particles at temperature $T = \sigma^2/2 k_B \alpha$, set by the FDT, where $k_B$ is the Boltzmann constant. Consequently, this substitution reduces Eq.~\ref{eq:generic_drho} to the standard Dean's equation for interacting Brownian particles at equilibrium \cite{dean1996langevin}.

We now linearize $\rho (\mathbf{x}, t)$ around a spatiotemporally constant density and derive the analytical static structure factor to get (see SI Sec.~III for derivation),
\begin{equation}
    \tilde{S}(\tilde{k}) = (1+2c) + [B(1+2c) - 2c] \,\tilde{k}^2,
    \label{eq:struc_fact}
\end{equation} 
where $B$ is a known system-dependent constant (SI Eq.~\ref{eq:B_pgd}). Eq.~\ref{eq:struc_fact} establishes the connection between the temporal noise correlation (quantified by $c$) and the long-range structure, and quantitatively predicts $S(k)$ for all noise correlations, without free parameters (Fig.~\ref{fig:figure2}b, SI Fig.~\ref{fig:figure1_si}). Additionally, since $\tilde{\delta \rho}^2 (\tilde{l} \to \infty) \propto \tilde{S}(\tilde{k} \to 0)$ \cite{torquato2018hyperuniform}, Eq.~\ref{eq:struc_fact} also quantitatively predicts the suppression of infinite wavelength density fluctuations (Fig.~\ref{fig:figure2}c inset).

Eq.~\ref{eq:struc_fact} has two competing terms. The first term $(1+2c)$ tends to make the long-range structure random ($\tilde{S}(\tilde{k}) \sim \text{const.}$), while the second term $[B(1+2c) - 2c] \, \tilde{k}^2$ tends to make it strongly hyperuniform ($\tilde{S}(\tilde{k}) \sim k^2$). This competition leads to the emergence of a normalized crossover length scale, given by,
\begin{equation}
    \tilde{l}_c = \sqrt{ B - \frac{2c}{ 1 + 2c } },
    \label{eq:l_c}
\end{equation}
which diverges for $c=-1/2$. Eq~\ref{eq:l_c} quantitatively predicts the crossover length scale for all noise correlations (Fig.~\ref{fig:figure2}b inset).

It is worth emphasizing that our theoretical framework remains valid across a broad range of parameter regimes, notably for arbitrary spatial dimension $d$, and for arbitrary noise correlation cutoff time $q$ (SI Sec.~III, Fig.~\ref{fig:figure1_si}).

The noise field $\boldsymbol{\psi}(\mathbf{x},t)$ appearing in Eq.~\ref{eq:generic_drho} is uncorrelated in space but exhibits temporal correlations. By contrast, in random-organizing systems where hyperuniformity emerges from local noisy interactions, the noise field at the continuum level is spatially correlated but uncorrelated in time \cite{anand2025emergent}. Consequently, the distinction between these two classes of systems is not limited to their microscopic dynamics but persists in their macroscopic (continuum) descriptions.

What is the regime of validity of our linearized fluctuating hydrodynamic theory used to predict the long-range structure? It is well established that such theories apply in the dense regime (i.e., at high particle volume fractions $\phi$) with soft overlap potentials \cite{illien2024dean, bon2025non, demery2014generalized, chavanis2008hamiltonian, anand2025emergent}. Consistently, we find that our theory does not accurately capture the numerical observations in the low-$\phi$ regime (SI Fig.~\ref{fig:figure2_si}).

\begin{figure*}[htpb!]
    \centering
    \includegraphics[width=\linewidth]{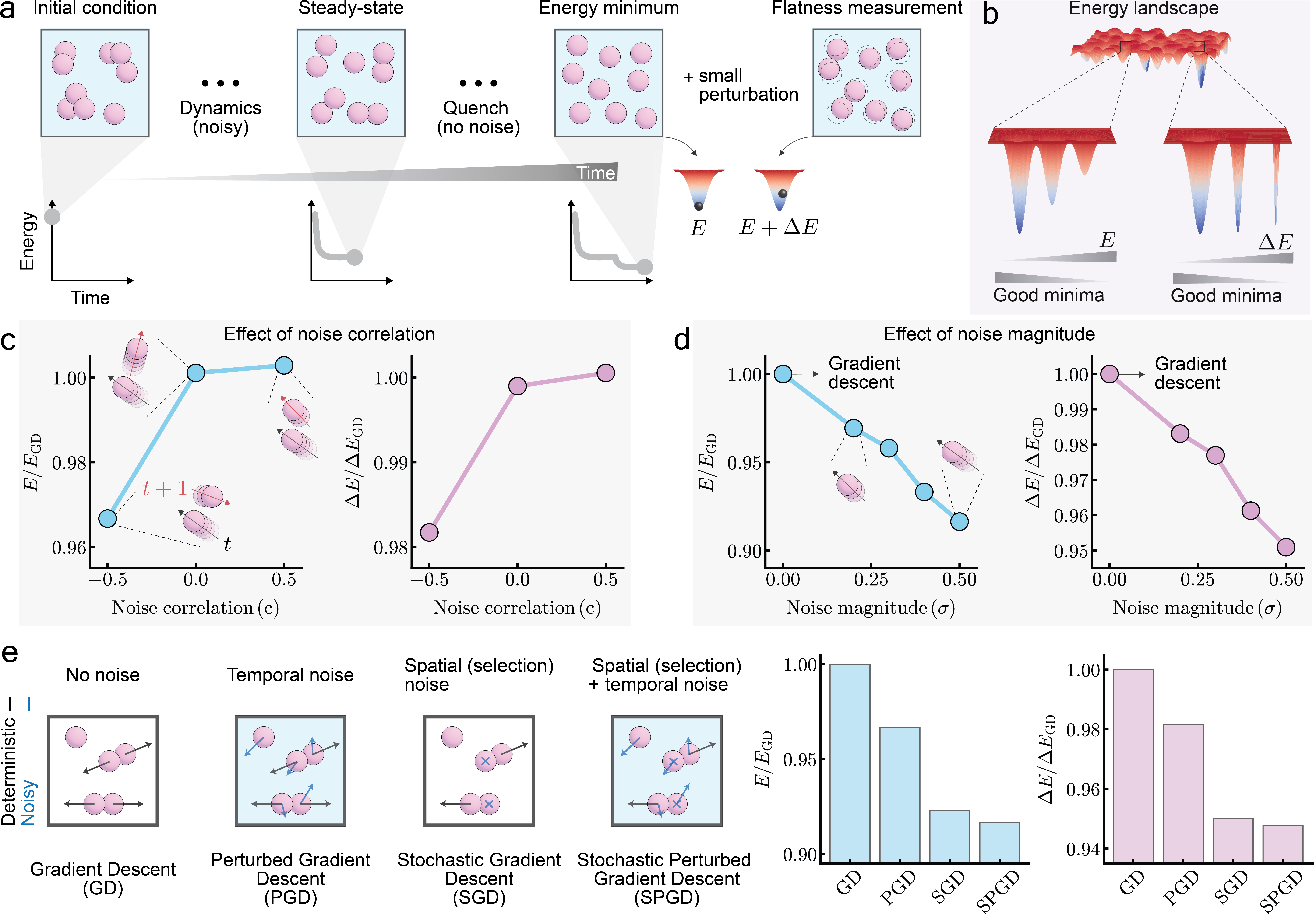}
    \caption{\small \textbf{Noise-driven exploration of energy landscape by perturbed gradient descent (PGD).} (a) Schematic of time evolution towards an energy minimum. Starting from a random initial condition, the system undergoes noisy dynamics according to Eq.~\ref{eq:discrete_pgd}, until it reaches a steady-state. The system is then quenched via (noiseless) gradient descent (GD), until it reaches an energy minimum. The system has an energy $E$ at the minimum and an energy $E + \Delta E$ after applying a small perturbation to the system. (b) Schematic of a generic energy landscape with multiple minima and maxima. ``Good'' minima are deeper and flatter, corresponding to low $E$ and $\Delta E$. (c) Normalized $E$ and $\Delta E$ versus noise correlation $c$ in particle simulations. Both quantities are normalized by the corresponding values from (noiseless) GD. Inset: Schematic of a single-particle motion without interactions. From time-step $t$ (black arrow) to $t+1$ (red arrow), for noise correlation: (i) $c = -0.5$, a particle typically reverses direction, (ii) $c = 0$, the motion is random, and (iii) $c=0.5$, the motion typically persists in the same direction. (d) Normalized $E$ and $\Delta E$ versus noise magnitude $\sigma$ in particle simulations. Both quantities are normalized by the corresponding values from (noiseless) GD ($\sigma = 0$). Insets: Schematic of a single-particle motion without interactions. (e) Comparison of (noiseless) GD and three different noisy dynamics (Left). Black arrows indicate deterministic pairwise interactions coming from the interaction potential, while blue arrows denote noise coming from the bath. Blue crosses indicate selection noise of stochastic gradient descent (SGD) (Eq.~\ref{eq:discrete_sgd_SI}, SI Sec.~IV). Stochastic perturbed gradient descent (SPGD) combines temporal noise from PGD with selection (or spatial) noise from SGD (Eq.~\ref{eq:discrete_spgd_SI}, SI Sec.~V). Light blue background denotes systems with temporal noise, originating from a non-equilibrium bath. Normalized $E$ and $\Delta E$ for the different dynamics in particle simulations (Right). Both quantities are normalized by the corresponding values from (noiseless) GD.
}
    \label{fig:figure3}
\end{figure*}

\subsection{Perturbed Gradient Descent}

PGD is a foundational algorithm in optimization and machine learning \cite{song2013stochastic, orvieto2022anticorrelated, orvieto2023explicit, choquette2023correlated, jin2021nonconvex}, widely used to train neural networks that use sensitive data, serving as the workhorse for differentially private learning \cite{ponomareva2023dp, dwork2006calibrating}. Differential privacy guarantees that the output of a neural network remains nearly unchanged whether or not a particular data point is included in the training set \cite{ponomareva2023dp}. This property has widespread practical importance, enabling applications such as training on medical records \cite{zhang2021adaptive, adnan2022federated}, fine-tuning large language models \cite{he2023exploring, hoory2021learning}, analyzing census data \cite{haney2021differentially}, and next-word prediction on mobile keyboards \cite{xu2023federated}. PGD achieves this privacy guarantee by deliberately injecting noise during training, thereby masking the contribution of any individual training datum. By contrast, standard stochastic gradient descent (SGD) \cite{bishop2023deep}, though ubiquitous in machine learning, does not provide differential privacy. In the literature, PGD is also often referred to as differentially private stochastic gradient descent (DP-SGD) \cite{beltran2024towards, choquette2023correlated, song2013stochastic, abadi2016deep}.

In neural networks, PGD updates the parameters (weights) according to Eq.~\ref{eq:discrete_pgd}, where particle positions are replaced by weights and the total energy $E = \sum_i \sum_{j \geq i} V_{ij}$ is replaced by an appropriate loss function \cite{song2013stochastic, orvieto2022anticorrelated, orvieto2023explicit, choquette2023correlated, jin2021nonconvex}.
At the end of optimization, the system settles into a minimum characterized by the value of the loss and the flatness of the landscape around it. Minima with both low loss and high flatness are typically considered ``good'', as they tend to generalize better on new data \cite{jastrzkebski2017three, xie2020diffusion, keskar2016large, huang2020understanding} (Fig.~\ref{fig:figure3}b).
Empirically, PGD dynamics in neural networks are found to reach ``good'' minima when (i) the noise becomes more anti-correlated, (ii) the noise magnitude increases (given the noise remains small enough), and (iii) the selection noise of SGD is also included in the dynamics \cite{orvieto2022anticorrelated, choquette2023correlated}. This naturally raises the question: Does PGD dynamics on particle-system energy landscapes similarly bias the system toward good minima, just as it does in the loss landscapes of neural networks?

We first allow the system to evolve until it reaches a steady-state (constant $E$) (Fig.~\ref{fig:figure3}a). To ensure proximity to an energy minimum, we then quench the system by removing the noise term in Eq.~\ref{eq:discrete_pgd} (Fig.~\ref{fig:figure3}a). Next, we measure the energy $E$ and the change in energy $\Delta E$ when a small noise is added to the system \cite{zhang2024absorbing, anand2025emergent} (Fig.~\ref{fig:figure3}a, Methods). Since $\Delta E \propto \text{Tr}(\mathbf{H})$, where $\mathbf{H}$ is the Hessian matrix of the system, $\Delta E$ serves as a measure of the flatness of the minimum \cite{zhang2024absorbing, jastrzkebski2017three}. Smaller $\Delta E$ values correspond to flatter minima (Fig.~\ref{fig:figure3}b).

We find that both $E$ and $\Delta E$ decrease as the noise becomes more anti-correlated and as its magnitude increases, demonstrating that temporal noise steers the system toward better minima (Figs.~\ref{fig:figure3}c, d). Notably, the presence of temporal noise yields better solutions than (noiseless) gradient descent ($\sigma = 0$) (Fig.~\ref{fig:figure3}d). 
Next, we incorporate the selection noise of SGD in addition to temporal noise, to get stochastic perturbed gradient descent (SPGD) (Eq.~\ref{eq:discrete_spgd_SI}, SI Sec.~V). In the context of particle systems, SGD corresponds to randomly selecting a subset of terms in the energy $E = \sum_i \sum_{j \geq i} V_{ij}$, and updating only the corresponding particle positions to minimize this partial energy (SI Sec.~IV) \cite{zhang2024absorbing, anand2025emergent}.
It was recently shown that the selection noise in SGD can be approximated by spatial noise in particle systems \cite{zhang2024absorbing}.  
We find that SPGD achieves better solutions than either PGD or SGD alone (Fig.~\ref{fig:figure3}e, Eqs.~\ref{eq:discrete_sgd_SI} and \ref{eq:discrete_spgd_SI}, SI Sec.~IV and V). 
Importantly, these results are robust: they hold independently of system parameters such as the learning rate $\alpha$, pairwise potential $V_{ij}$, volume fraction $\phi$, noise correlation cutoff parameter $q$, and the spatial dimension $d$ (SI Fig.~\ref{fig:figure3_si}).

\subsection{Discussion}

Our results, combining simulations and theory, establish short-memory temporal noise as a novel mechanism for noise-driven self-organization. We demonstrate that temporal correlations can induce self-organization at large length scales, and uncover the microscopic mechanism underlying this emergent behavior. Finally, we show that the temporal noise–driven navigation of particle system energy landscapes closely parallel the loss landscapes of neural networks, revealing striking connections between these disparate systems.

Is the emergence of long-range structure universal to other short-memory noises? Because MA($q$) noise can model arbitrary correlation structures, we approximate the widely studied exponentially correlated noise (the Ornstein–Uhlenbeck process) by tuning $q$ and the coefficients $\{\theta_k\}_{k=0}^{q}$. We find that the long-range structure is random ($S(k \to 0) = \text{const.}$)---thus demonstrating that anti-correlations are essential for large-scale suppression of density fluctuations (SI Fig.~\ref{fig:figure4_si}).

It has been recently shown that SGD dynamics on interacting particle systems can suppress density fluctuations and generate hyperuniformity when the selection noise, which can be approximated by spatial noise, is anti-correlated \cite{anand2025emergent}. This raises the question of how the combined presence of selection (spatial) and temporal noise---i.e., SPGD dynamics---affect the long-range structure. 
We perform particle simulations and find that achieving hyperuniformity requires strong anti-correlation in both spatial and temporal noise simultaneously (SI Fig.~\ref{fig:figure5_si}, SI Sec.~V). Further, we extend our fluctuating hydrodynamic theory to incorporate spatiotemporal noise, and show that it quantitatively captures the numerical observations (SI Eq.~\ref{eq:sk_spgd_simplified_MA1_norm}, SI Sec.~V).

Strong hyperuniformity where the structure factor follows the scaling $S(k) \sim k^2$ is typically observed in systems that conserve the local center of mass, either through deterministic \cite{lei2019hydrodynamics, lei2019nonequilibrium, kuroda2023microscopic} or spatially noisy dynamics \cite{hexner2017noise, galliano2023two, anand2025emergent, maire2025hyperuniformity}. More recently, it has been shown that systems with long-memory temporal noise can also exhibit hyperuniformity \cite{ikeda2024harmonic, ikeda2023correlated, ikeda2024continuous}. In contrast, our results demonstrate that strong hyperuniformity can arise without requiring either local center of mass conservation or long-memory temporal noise---thus opening a new direction for both fundamental studies and practical applications of temporal noise-driven hyperuniformity.

Disordered hyperuniform structures have diverse applications, ranging from optical uses such as isotropic photonic bandgaps \cite{man2013photonic, shih2024fast} and transparency \cite{leseur2016high} to materials applications such as high absorption rate solar cells \cite{tavakoli2022over}. 
Further, in $2d$, strong hyperuniformity is essential for long-range order \cite{galliano2023two}. Therefore, developing new and robust routes for assembling hyperuniform structures is highly desirable. 
While short-range spatial noise in random-organizing systems leads to strong hyperuniformity, the noise originates from the particles themselves---i.e., the particles are \textit{active} \cite{anand2025emergent, hexner2017noise}. 
In contrast, temporally correlated noise offers a route to assemble \textit{passive} particles (e.g., colloids) into hyperuniform structures by simply immersing them in an active bath, whose properties can be tuned to achieve the desired long-range structure.

Our study demonstrates that temporally correlated optimization algorithms (e.g., PGD) exhibit a similar ability to reach good minima in both interacting particle systems and deep neural networks, much like spatially correlated algorithms (e.g., SGD \cite{zhang2024absorbing, anand2025emergent}). Because particle systems are more tractable, both numerically and analytically, they provide a useful playground for developing novel optimization algorithms, with potential applications in high-dimensional optimization and machine learning.



\bibliographystyle{apsrev4-2}
\bibliography{ref.bib}

\section{Acknowledgements}

We thank Mathias Casiulis for careful reading and comments on the manuscript and Mathias Casiulis, Flaviano Morone, and Shivang Rawat for insightful discussions. This work was supported by the National Science Foundation grant IIS-2226387, National Institute of Health under award number R01MH137669, Simons Center for Computational Physical Chemistry, and in part by the NYU IT High Performance Computing resources, services, and staff expertise.

\section{Author contributions}

S.A. conceived the project. S.A. and G.Z. performed simulations. S.A. developed the theory. S.A., G.Z, and S.M. analyzed data and wrote the manuscript. S.M. acquired funding and supervised the research.

\section{Methods}

Our system consisted of $N$ hyperspherical particles, each with radius $R$, confined within a $d$-dimensional hypercubic box of side length $L$ under periodic boundary conditions. We defined the unit of length, time, and energy as $2R$, $\tau$, and $\mathcal{E}$, respectively, where $\mathcal{E}=1$ sets the characteristic energy scale of the interaction potential given by Eq.~\ref{eq:v_ji_family}, and $\tau = 1$ is the time scale corresponding to a discrete time-step. The particle radius was fixed at $R = 1$, and the particle volume fraction $\phi = N V_s / V_c$ was tuned by varying the box size $L$. $V_c$ and $V_s$ denote volumes of the $d$-dimensional cube of side length $L$ and sphere of radius $R$, respectively. At $t=0$, particles were placed randomly within the simulation box. Simulations were run until the system reached a steady-state, defined as the regime where the energy $E = \sum_i \sum_{j \geq i} V_{ij}$ fluctuated around a constant mean value. 

The particle dynamics were updated according to Eq.~\ref{eq:discrete_pgd}, with $ V_{ij}$ given by a family of short-range, purely repulsive potentials,
\begin{equation}
    V_{ij}(r) =     
    \begin{cases}
      \frac{\mathcal{E}}{p}\left(1 - \frac{r_{ij}}{2R}\right)^p, & \text{if}\ 0 < r_{ij} < 2R, \\
      0, & \text{otherwise},
    \end{cases}  
    \label{eq:v_ji_family}
\end{equation}
where $r_{ij} = |\mathbf{x}_j - \mathbf{x}_i|$, $\mathcal{E}$ specifies the characteristic energy scale, and $p$ controls the stiffness of the potential. For systems subject to MA($1$) noise, the dynamics depend on six control parameters: the learning rate $\alpha$, potential stiffness $p$, noise magnitude $\sigma$, noise correlation coefficient $c$, volume fraction $\phi$, and the spatial dimension $d$.
\\

\textbf{Density fluctuations and structure factor.} The structure factor of a particle system is given as $S(\mathbf{k}) = |\hat{\rho}(\mathbf{k})|^2/N$, where $\rho(\mathbf{x}) = \sum_{i=1}^{N} \delta (\mathbf{x} - \mathbf{x}_i)$. The spatial Fourier transform of a generic function $f(\mathbf{x})$ is given by $\hat{f}(\mathbf{k}) = \int f(\mathbf{x}) e^{-i \mathbf{k} \cdot \mathbf{x}} d \mathbf{x}$. $S(\mathbf{k})$ was computed using the nonuniform fast Fourier transform \cite{barnett2019parallel, barnett2021aliasing}. $S(\mathbf{k})$ was then radially averaged to obtain the radial structure factor $S(k)$.

The variance of number density $\delta \rho^2(l)$ in a hyperspherical observation window of diameter $l$ was obtained through its exact relation to $S(k)$ \cite{torquato2018hyperuniform}, 
\begin{equation}
    \delta \rho^2(l) = \frac{\bar{\rho} d 2^d \Gamma(1 + \frac{d}{2})}{( \sqrt{\pi} l)^d} \int_0^{\infty} \frac{1}{k} S(k) \left[ J_{d/2} \left( \frac{kl}{2}\right) \right]^2 \, dk,
    \label{sk_den_fluc_rel}
\end{equation}
where $\bar{\rho} = N/V_c$, $J$ is the Bessel function of the first kind, and $\Gamma$ denotes the Gamma function. The integral in Eq.~\ref{sk_den_fluc_rel} was numerically evaluated using Simpson's rule, based on the measured $S(k)$.

For the results shown in Fig.~\ref{fig:figure2}, parameters were set as $N=100000$, $\alpha = 1.5$, $p = 1.5$, $\sigma = 1.0$, $\phi = 1.0$, $d = 2$, and $c \in [-0.5, 0]$. All observables were computed at steady-state and all quantities were averaged over $100$ steady-state configurations.
\\

\textbf{Energy and flatness near the minimum.} Beginning from a random initial condition, the system first evolved to a steady state, characterized by a constant value of the total energy $E$. To ensure proximity to an energy minimum, the system was quenched by removing the noise and evolved under (noiseless) gradient descent dynamics (Eq.~\ref{eq:discrete_pgd} without the noise term). The system was evolved until the energy no longer changed, and this final value was recorded as the energy at the minimum. To probe the flatness of the energy minimum, the final configuration $\mathbf{X} \equiv \{ \mathbf{x}_1, \mathbf{x}_2,...,\mathbf{x}_N \}$ having energy $E$ was perturbed by adding an independent Gaussian noise $\mathbf{N}_i(\mathbf{0}, \beta^2 \mathbf{I})$ to the position of every particle, yielding the perturbed configuration $\mathbf{X} + \Delta \mathbf{X} \equiv \{ \mathbf{x}_1+\mathbf{N}_1, \mathbf{x}_2+\mathbf{N}_2,...,\mathbf{x}_N+\mathbf{N}_N \}$ with energy $E + \Delta E$.  

For the results shown in Fig.~\ref{fig:figure3}, the base parameters were set as $N=10000$, $\alpha = 0.1$, $p = 1.5$, $\phi = 1.0$, $d = 2$, and $\beta = 0.01$. 
For results in Fig.~\ref{fig:figure3}c, additional parameters were set as $\sigma = 0.25$. 
For results in Fig.~\ref{fig:figure3}d, additional parameters were set as $c = -0.5$.
SGD and SPGD dynamics required two additional parameters: the batch fraction $b_f$ and the Pearson correlation coefficient of the selection (spatial) noise $c^s$ (SI Eqs.~\ref{eq:discrete_sgd_SI} and \ref{eq:discrete_spgd_SI}, SI Sec.~IV and V). 
For results in Fig.~\ref{fig:figure3}e, additional parameters for PGD were set as $\sigma = 0.25$, $c = -0.5$, for SGD, $b_f = 0.5$, $c^s = 0$, and for SPGD, $b_f = 0.5$, $c^s = 0$, $\sigma = 0.25$, $c = -0.5$.
All results were averaged over $100$ independent simulations and $5000$ independent realizations of noise $\mathbf{N}_i$.

\end{document}


\preprint{APS/123-QED}

\title{\textbf{\LARGE{Supplementary Information}} \\ Spatial self-organization driven by temporal noise}

\author{Satyam Anand}
\email{sa7483@nyu.edu}
\affiliation{Courant Institute of Mathematical Sciences, New York University, New York, NY 10003, USA}
\affiliation{\mbox{Center for Soft Matter Research, Department of Physics, New York University, New York, NY 10003, USA}}

\author{Guanming Zhang}
\affiliation{\mbox{Center for Soft Matter Research, Department of Physics, New York University, New York, NY 10003, USA}}
\affiliation{\mbox{Simons Center for Computational Physical Chemistry, Department of Chemistry, New York University, New York, NY 10003, USA}}

\author{Stefano Martiniani}
\email{sm7683@nyu.edu}
\affiliation{Courant Institute of Mathematical Sciences, New York University, New York, NY 10003, USA}
\affiliation{\mbox{Center for Soft Matter Research, Department of Physics, New York University, New York, NY 10003, USA}}
\affiliation{\mbox{Simons Center for Computational Physical Chemistry, Department of Chemistry, New York University, New York, NY 10003, USA}}
\affiliation{\mbox{Center for Neural Science, New York University, New York, NY 10003, USA}}

\maketitle 


\section{Continuous-time approximation of discrete-time dynamics}

We consider particles interacting via a pairwise potential immersed in a non-equilibrium bath. 
The position $\mathbf{x}_{i}^{m}$ of particle $i$ at time-step $m$ evolves as,
\begin{equation}
  \mathbf{x}_{i}^{m} = \mathbf{x}_{i}^{m-1} - \underbrace{ \alpha \sum_{j \in \Gamma_i^{m-1}} \nabla_{\mathbf{x}_{i}} V_{ji}^{m-1} }_{\text{deterministic term}} + \underbrace{ \dfrac{\sigma}{\Theta} \sum_{k=0}^{q} \theta_k \boldsymbol{\xi}_i^{m-k} }_{\text{noise term}},
\label{eq:discrete_pgd_SI}
\end{equation}
where $\Theta = \sqrt{\sum_{k=0}^{q} \theta_k^2}$, $\alpha$ is the learning rate with units of $\text{length}/\text{force}$, $V_{ji}^{m-1} = V(|\mathbf{x}^{m-1}_{j} - \mathbf{x}^{m-1}_{i}|)$ is a pairwise interaction potential between particles $i$ and $j$, and $\Gamma_i^{m-1} = \{j \mid |\mathbf{x}_j^{m-1} - \mathbf{x}_i^{m-1}| < 2R,\ j \neq i \}$ is the set containing all particles that overlap with particle $i$ at time-step $m-1$. $R$ is the particle radius. $\boldsymbol{\xi}_i^{m}$ is a standard Gaussian, white noise with $\left< {\xi}_{i, \alpha}^{m} \right> = 0$, and $\left< {\xi}_{i, \alpha}^{m} {\xi}_{j, \beta}^{n} \right> = \delta_{ij} \delta_{\alpha \beta} \delta^{mn}$. $\sigma$ has the units of $\text{length}$ and controls the noise magnitude. $q$ is the number of time-steps up to which the noise has non-zero correlations. Finally, $\theta_0 = 1$, and $\{\theta_k\}_{k=1}^{q} \in [-1, 1]$ are scalar parameters that weight noise from the previous $q$ time steps. While $V_{ji}$ can be any potential (short- or long-range), we consider short-range, repulsive, cutoff potentials given by,
\begin{equation}
    V_{ij}(r) =     
    \begin{cases}
      \dfrac{\mathcal{E}}{p}\left(1 - \dfrac{r_{ij}}{2R}\right)^p, & \text{if}\ 0 < r_{ij} < 2R, \\
      0, & \text{otherwise},
    \end{cases}  
    \label{eq:v_ji_family_SI}
\end{equation}
where $r_{ij} = |\mathbf{x}_j - \mathbf{x}_i|$, $\mathcal{E}$ specifies the characteristic energy scale, and $p$ controls the stiffness of the potential.

The noise in Eq.~\ref{eq:discrete_pgd_SI} is called the moving-average $q$ (MA($q$)) noise \cite{shumway2006time}. The cross-covariance of the complete noise vector $\boldsymbol{\zeta}^m_{i} = \dfrac{\sigma}{\sqrt{\sum_{k=0}^{q} \theta_k^2}} \sum_{k=0}^{q} \theta_k \boldsymbol{\xi}_i^{m-k}$ is
\begin{equation}
 \text{cov}[ \zeta_{i, \alpha}^m \zeta_{j, \beta}^n] =  \begin{cases}
  \delta_{ij} \delta_{\alpha \beta} \dfrac{\sigma^2}{\sum_{k=0}^{q} \theta_k^2} \sum_{k=0}^{q-s} \theta_k \theta_{k+s}, & \text{if}\ 0 \le s \le q, \\
  0_{\alpha \beta}, & \text{if}\ s > q,  
  \end{cases}
 \label{eq:cov_dis}
\end{equation}
where $s = |m-n|$. The Pearson correlation coefficient between the total noise $\zeta_{i, \alpha}^m$ and $\zeta_{i, \alpha}^{m-h}$ is then,
\begin{equation}
  c_h =  \begin{cases}
  \dfrac{1}{\sum_{j=0}^q \theta_j^2} \sum_{k=0}^{q-h} \theta_k \theta_{k+h}, & \text{if}\ 1 \leq h \leq q, \\
  0, & \text{if}\ h > q.  
  \end{cases}
 \label{eq:corr_coeff}
\end{equation}

It is instructive to consider bounds on the sum $\sum_{h=1}^{q} c_h$. Using Eq.~\ref{eq:corr_coeff}, we can write,
\begin{equation}
\sum_{h=1}^{q} c_h = \frac{1}{\sum_{j=0}^{q} \theta_j^2} \sum_{0 \le j < k \le q} \theta_j \theta_k = \frac{1}{2\sum_{j=0}^{q} \theta_j^2}
\left[
\left( \sum_{j=0}^{q} \theta_j \right)^2
- \sum_{j=0}^{q} \theta_j^2
\right] = \frac{1}{2} \left( \frac{\left( \sum_{j=0}^{q} \theta_j \right)^2}{\sum_{j=0}^{q} \theta_j^2} - 1 \right).
 \label{eq:sum_c}
\end{equation}
We now use Sedrakyan's inequality (a direct consequence of the Cauchy–Schwarz inequality) to get,
\begin{equation}
0 \leq \frac{\left( \sum_{j=0}^{q} \theta_j \right)^2}{\sum_{j=0}^{q} \theta_j^2} \leq q+1.
 \label{eq:csz_ineq}
\end{equation}
Combining Eq.~\ref{eq:sum_c} with Eq.~\ref{eq:csz_ineq}, we get,
\begin{equation}
-\frac{1}{2} \leq \sum_{s=1}^{q} c_s \leq \frac{q}{2}.
\label{eq:sum_c_bounds}
\end{equation}
The lower bound in Eq.~\ref{eq:sum_c_bounds} is attained when $\sum_{j=0}^{q} \theta_j = 0$, while the upper bound is attained when all coefficients are equal, $\theta_j = 1 \, \forall \, j=0,\ldots,q$. As a direct consequence of Eq.~\ref{eq:sum_c_bounds}, for the special case of MA($1$) noise, the lag-$1$ noise correlation coefficient satisfies $c_1 \equiv c \in [-1/2, 1/2]$.

We now aim to write a continuous-time approximation of the discrete-time dynamics given by Eq.~\ref{eq:discrete_pgd_SI}. The standard approach is to use the framework of stochastic modified equations, in which the discrete-time dynamics is interpreted as the Euler-Maruyama discretization of a continuous-time stochastic differential equation (SDE) \cite{li2017stochastic, zhang2024absorbing, anand2025emergent}. However, this framework applies only to white noise, whereas the noise in Eq.~\ref{eq:discrete_pgd_SI} is colored. Here, we take a different approach: we compute the power spectral density (PSD) of the MA($q$) noise and then construct a continuous-time noise with a matching PSD. 

The PSD of a discrete process is given by,
\begin{equation}
 P(\omega) = \sum_{s=-\infty}^{\infty} R_{\omega\omega}(s\tau) e^{-is\omega\tau}.
 \label{eq:psd_def_general}
\end{equation}
where $R_{xx}$ is the autocorrelation function. For the MA($q$) noise, using $R_{xx}(-s\tau) = R_{xx}(s\tau)$, we get the PSD as,
\begin{align}
 P(\omega) 
 &= R_{\omega\omega}(0) + 2 \sum_{s=1}^{q} R_{xx}(s\tau) \cos(s\omega\tau)  \nonumber \\
 &= \dfrac{\sigma^2}{\sum_{k=0}^{q} \theta_k^2} \left[ \sum_{k=0}^{q} \theta_k^2 + 2 \sum_{s=1}^{q} \left(\sum_{k=0}^{q-s} \theta_k \theta_{k+s}\right) \cos(s\omega\tau) \right],
 \label{psd_dis}
\end{align}
where $\tau$ is the characteristic time elapsed in a discrete time-step. We now write an overdamped Langevin equation for the
evolution of the position of particle $i$,
\begin{equation}
 \dfrac{d\mathbf{x}_{i}(t)}{dt}
 = \underbrace{ - \dfrac{1}{\gamma} \sum_{j = 1}^N \nabla_{\mathbf{x}_i} V_{ji}}_{\text{deterministic term}} 
 + \underbrace{ \sqrt{2D} \ \boldsymbol{\eta}_{i}(t)}_{\text{noise term}} 
 \label{eq:generic_sde_pgd} 
\end{equation}
where $\gamma = \tau/\alpha$ is the friction coefficient, and $D = \sigma^2/2 \tau$ is the diffusion constant. $\boldsymbol{\eta}_{i}(t)$ is a Gaussian colored noise with mean $\left< {\eta}_{i, \alpha}(t) \right> = 0$, and the covariance matrix given by,
\begin{align}
\left< {\eta}_{i,\alpha} (t) \, {\eta}_{j,\beta} (t') \right> 
&= \dfrac{\delta_{ij} \delta_{\alpha \beta}}{\sum_{k=0}^{q} \theta_k^2} \left\{ \left( \sum_{k=0}^{q} \theta_k^2 \right) \delta(t-t') + \sum_{s=1}^{q} \left( \sum_{k=0}^{q-s} \theta_k \theta_{k+s} \right) \left[ \delta(t-t'-s\tau) + \delta(t-t'+s\tau) \right] \right\} \nonumber \\
&= \dfrac{\delta_{ij} \delta_{\alpha \beta}}{\sum_{k=0}^{q} \theta_k^2} \left\{ \left( \sum_{k=0}^{q} \theta_k^2 \right) \delta(t-t') + \sum_{s=1}^{q} \left( \sum_{k=0}^{q-s} \theta_k \theta_{k+s} \right)  \delta(|t-t'|-s\tau) \right\},
\label{eq:cov_eta_continuous}
\end{align}
where the second step follows since $\tau > 0$. From Eq.~\ref{eq:cov_eta_continuous}, we see that the noise is stationary. The PSD of Eq.~\ref{eq:cov_eta_continuous} is identical to Eq.~\ref{psd_dis}. Thus Eq.~\ref{eq:generic_sde_pgd} is the continuous-time approximation of the discrete-time dynamics given by Eq.~\ref{eq:discrete_pgd_SI}.

Note that when $\theta_k = 0 \ \forall \ k \geq 1$, Eq.~\ref{eq:generic_sde_pgd} reduces to an equilibrium system of interacting Brownian particles (the noise $\boldsymbol{\eta}_{i}(t)$ becomes white), at temperature $T = \sigma^2/2 k_B \alpha$, set by the fluctuation-dissipation theorem, where $k_B$ is the Boltzmann constant.

Before proceeding, we note that coarse-graining our system to obtain a fluctuating hydrodynamic theory requires the use of It\^{o} calculus. Since It\^{o} calculus applies to processes that are semi-martingales \cite{klebaner2012introduction, Iyer}, we first verify whether the integrated noise $\boldsymbol{\eta}_i(t)$ qualifies as a semi-martingale.

\section{Validity of It\^{o} calculus}

A stochastic process is a semi-martingale if it can be decomposed into the sum of a local martingale (the ``rough'' stochastic part) and a finite variation process (the ``smooth'' predictable part) \cite{klebaner2012introduction, Iyer}. We first decompose the noise $\boldsymbol{\eta}_i(t)$ into two statistically independent parts as (Eq.~\ref{eq:cov_eta_continuous}),
\begin{equation}
\boldsymbol{\eta}_i(t) = \boldsymbol{\eta}_{i}^W(t) + \boldsymbol{\eta}_{i}^C(t),
\label{eq:decomp_noise}
\end{equation}
where $\boldsymbol{\eta}_{i}^W(t)$ is a standard Gaussian white noise with covariance $\left< \eta_{i,\alpha}^W (t) \, \eta_{j,\beta}^W (t') \right> = \delta_{ij} \delta_{\alpha \beta} \delta(t-t')$, and $\boldsymbol{\eta}_{i}^C(t)$ is the colored component with covariance $\left< \eta_{i,\alpha}^C (t) \, \eta_{j,\beta}^C (t') \right> = (1/ \sum_{k=0}^{q} \theta_k^2) \sum_{s=1}^{q} \left( \sum_{k=0}^{q-s} \theta_k \theta_{k+s} \right) \delta_{ij} \delta_{\alpha \beta} \delta(|t-t'| - s \tau)$. Integrating this decomposition gives
\begin{equation}
\mathbf{Z}_i(t) = \int_0^t \boldsymbol{\eta}_i(u) du = \underbrace{\int_0^t \boldsymbol{\eta}_{i}^W(u) du}_{\mathbf{E}(t)} + \underbrace{\int_0^t \boldsymbol{\eta}_{i}^C(u) du}_{\mathbf{F}(t)}.
\label{eq:integrated_noise}
\end{equation}
The integral of Gaussian white noise, $\mathbf{E}(t)$, is a multi-dimensional Wiener process, and thus the canonical example of a martingale. We now show that the second term, $\mathbf{F}(t)$, is a finite variation process.

If the quadratic variation of a stochastic process is zero, it is of finite variation \cite{klebaner2012introduction, Iyer}. The quadratic variation of $F_{\alpha}(t)$ is given by,
\begin{align}
d[F_{\alpha}, F_{\beta}]_t 
&= \left< dF_{\alpha}(t) \, dF_{\beta}(t) \right> \nonumber \\
&= \left< \left( \int_t^{t+dt} \eta_{i, \alpha}^C(u) du \right) \left( \int_t^{t+dt} \eta_{j, \beta}^C(u') du' \right) \right> \nonumber \\
&= \int_t^{t+dt} \int_t^{t+dt} \left< \eta_{i, \alpha}^C(u) \eta_{j, \beta}^C(u') \right> du \, du' \nonumber \\
&= \dfrac{\delta_{ij} \delta_{\alpha \beta}}{\sum_{k=0}^{q} \theta_k^2} \int_t^{t+dt} \int_t^{t+dt} \sum_{s=1}^{q} \left( \sum_{k=0}^{q-s} \theta_k \theta_{k+s} \right) \delta(|u-u'| - s \tau) \, du \, du'.
\label{eq:quad_var_F}
\end{align}
The domain of this double integral is an infinitesimal square in the $u-u'$ plane with corners at $(t,t)$ and $(t+dt, t+dt)$. For the delta function $\delta(|u-u'| - \tau)$ to be non-zero, the condition $|u-u'| = \tau$ must be satisfied within this domain. However, for any two points $u, u' \in [t, t+dt]$, the maximum possible difference is $|u-u'|_{\text{max}} = dt$. For any finite time lag $\tau > 0$, the condition $|u-u'| = \tau$ can never be met inside the infinitesimal integration domain, since we consider the limit $dt \to 0$. Thus, the argument of the delta function is never zero. Consequently, this gives,
\begin{equation}
d[F_{\alpha}, F_{\beta}]_t = 0.
\label{eq:quad_var_F_2}
\end{equation}
Since the quadratic variation of the process $\mathbf{F}(t)$ is zero, it is a finite variation process \cite{klebaner2012introduction, Iyer}. Thus, $\mathbf{Z}_i(t) = \mathbf{E}(t) + \mathbf{F}(t)$ is the sum of a martingale and a finite variation process, and by definition, a semi-martingale. Hence, It\^{o} calculus can be applied to the noise $\boldsymbol{\eta}_i(t)$.

It\^{o}'s lemma requires the knowledge of the quadratic variation of the full noise $\mathbf{Z}_i(t)$, which is given by,
\begin{align}
d[Z_{i,\alpha}, Z_{j,\beta}]_t 
&= d[E_{\alpha}+F_{\alpha}, E_{\beta}+F_{\beta}]_t \nonumber \\
&= d[E_{\alpha}, E_{\beta}]_t + d[E_{\alpha}, F_{\beta}]_t + d[F_{\alpha}, E_{\beta}]_t + d[F_{\alpha}, F_{\beta}]_t \nonumber \\
&= d[E_{\alpha}, E_{\beta}]_t \nonumber \\
&= \delta_{ij}\delta_{\alpha\beta} \, dt,
\end{align}
where we have used Eq.~\ref{eq:quad_var_F_2}, and the fact that since $\mathbf{F}(t)$ is of finite variation, all cross-terms are zero. Thus, despite the presence of temporal correlations in $\boldsymbol{\eta}_i(t)$ (Eq.~\ref{eq:cov_eta_continuous}), the quadratic variation of $\mathbf{Z}_{i}$ is identical to that of the Wiener process, allowing It\^{o}'s lemma to be applied without modification.

\section{Fluctuating hydrodynamics}

Taking the continuous-time SDE (Eq.~\ref{eq:generic_sde_pgd}) as our starting point, we now aim to derive a coarse-grained continuum description of our system. 
We employ Dean’s method for this purpose, which has been used to systematically coarse-grain systems with additive \cite{dean1996langevin}, multiplicative \cite{bertin2013mesoscopic, solon2015active, donev2014dynamic, anand2025emergent}, and colored \cite{russo2020memory, feng2021effective} noise. 

The density function associated with a single particle $\rho_{i} (\mathbf{x}, t)$ is defined as,
\begin{equation}
 \rho_{i} (\mathbf{x}, t)
 = \delta(\mathbf{x}_i (t) - \mathbf{x}). 
 \label{eq:density_1_part}
\end{equation}
The global density $\rho (\mathbf{x}, t)$ is then given by,
\begin{equation}
 \rho (\mathbf{x}, t)
 = \sum_{i=1}^{N} \delta(\mathbf{x}_i (t) - \mathbf{x}). 
 \label{eq:density_global}
\end{equation}
Let $f(\mathbf{x}_i (t))$ be an arbitrary function defined on the coordinate space of the system. Using the definition of $\rho_{i} (\mathbf{x}, t)$, we can then write,
\begin{equation}
 f(\mathbf{x}_i (t))
 = \int f(\mathbf{x}) \rho_{i} (\mathbf{x}, t) d\mathbf{x}.
 \label{eq:arbt_func_f}
\end{equation}
We now expand Eq.~\ref{eq:generic_sde_pgd} using It\^{o} calculus to get,
\begin{align}
 \dfrac{df(\mathbf{x}_i (t))}{dt} 
 &= - \dfrac{1}{\gamma} \sum_{j} \partial_{{x}_i, \alpha} V(\mathbf{x}_i, \mathbf{x}_j) \, \partial_{{x}_i, \alpha} f(\mathbf{x}_i (t)) + \sqrt{2 D } \ \eta_{i, \alpha}(t)  \partial_{{x}_i, \alpha} f(\mathbf{x}_i (t)) + \dfrac{1}{2} 2D \ \partial_{{x}_i, \alpha} \partial_{{x}_i, \alpha} f(\mathbf{x}_i (t)).
 \label{eq:df_dt_1}
\end{align}
Using Eq.~\ref{eq:arbt_func_f}, and noting that $\partial_{{x}_i, \alpha} f(\mathbf{x}_{i}) = \int \rho_{i}(\mathbf{x},t) \partial_{{x}, \alpha} f(\mathbf{x}) d\mathbf{x}$, and $\partial_{{x}_i, \alpha} \partial_{{x}_i, \gamma} f(\mathbf{x}_{i}) = \int \rho_{i}(\mathbf{x},t) \partial_{{x}, \alpha} \partial_{{x}, \gamma} f(\mathbf{x}) d\mathbf{x}$, we get,
\begin{align}
 \dfrac{df(\mathbf{x}_i (t))}{dt} 
 &= \int d\mathbf{x} \, \rho_{i}(\mathbf{x}, t) \Bigg[ - \dfrac{1}{\gamma} \sum_{j} \partial_{{x}, \alpha} V(\mathbf{x}, \mathbf{x}_j) \, \partial_{{x}, \alpha} f(\mathbf{x}) + \sqrt{2D} \ \eta_{i,\alpha}(t) \, \partial_{{x}, \alpha} f(\mathbf{x}) + D \ \partial_{{x}, \alpha} \partial_{{x}, \alpha} f(\mathbf{x}) \Bigg]. 
 \label{eq:df_dt_2}
\end{align}
Applying integration by parts on Eq.~\ref{eq:df_dt_2} and assuming $f(|\mathbf{x}|\to \infty)=0$ leads to,
\begin{align}
 \dfrac{df(\mathbf{x}_i (t))}{dt} 
 &= \int d\mathbf{x} \, f(\mathbf{x}) \Bigg[ \dfrac{1}{\gamma} \partial_{\alpha} \left( \rho_{i}(\mathbf{x},t) \sum_{j} \partial_{\alpha} V(\mathbf{x}, \mathbf{x}_j) \right) - \sqrt{2D} \ \partial_{\alpha} \left( \rho_{i}(\mathbf{x},t)   \ \eta_{i, \alpha}(t) \right) + D \partial_{\alpha} \partial_{\alpha} \left( \rho_{i}(\mathbf{x},t) \right) \Bigg],
 \label{eq:df_dt_3}
\end{align}
where $\partial_{\alpha} = \partial_{{x}, \alpha}$ henceforth. From Eq.~\ref{eq:arbt_func_f}, we also have,
\begin{equation}
 \dfrac{df(\mathbf{x}_i (t))}{dt} 
 = \int d\mathbf{x} \dfrac{\partial \rho_{i}(\mathbf{x}, t)}{\partial t} f(\mathbf{x}). 
 \label{eq:df_dt_4}
\end{equation}
Comparing Eqs.~\ref{eq:df_dt_3} and \ref{eq:df_dt_4}, we get,
\begin{equation}
 \dfrac{\partial \rho_{i}(\mathbf{x}, t)}{\partial t} 
 = \dfrac{1}{\gamma} \partial_{\alpha} \left( \rho_{i}(\mathbf{x},t) \sum_{j} \partial_{\alpha} V(\mathbf{x}, \mathbf{x}_j) \right) - \sqrt{2D} \ \partial_{\alpha} \left( \rho_{i}(\mathbf{x},t) \ \eta_{i,\alpha}(t) \right) + 
 D \partial_{\alpha} \partial_{\alpha} \left( \rho_{i}(\mathbf{x},t) \right).
 \label{eq:drho_i_dt}
\end{equation}
We now sum Eq.~\ref{eq:drho_i_dt} over $i = 1$ to $N$ to get,
\begin{align}
 \dfrac{\partial \rho(\mathbf{x}, t)}{\partial t} 
 &= \sum_{i} \dfrac{1}{\gamma} \partial_{\alpha} \left( \rho_{i}(\mathbf{x},t) \sum_{j} \partial_{\alpha} V(\mathbf{x}, \mathbf{x}_j) \right) 
  - \sum_{i} \sqrt{ 2D} \partial_{\alpha} \left( \rho_{i}(\mathbf{x},t) \ \eta_{i, \alpha}(t) \right) + 
 \sum_{i} D \partial_{\alpha} \partial_{\alpha} \left( \rho_{i}(\mathbf{x},t) \right).
 \label{eq:drho_dt_1}
\end{align}
Consider the first term on the right hand side (RHS) of Eq.~\ref{eq:drho_dt_1},
\begin{align}
 \sum_{i} \dfrac{1}{\gamma} \partial_{\alpha} \left( \rho_{i}(\mathbf{x},t) \sum_{j} \partial_{\alpha} V(\mathbf{x}, \mathbf{x}_j) \right) 
 &= \dfrac{1}{\gamma} \partial_{\alpha} \left( \sum_i \rho_{i}(\mathbf{x},t) \sum_{j} \partial_{\alpha} V(\mathbf{x}, \mathbf{x}_j) \right) \nonumber \\ 
 &= \dfrac{1}{\gamma} \partial_{\alpha} \left( \rho(\mathbf{x},t) \int d\mathbf{y} \sum_{j}  \rho_j(\mathbf{y},t) \partial_{\alpha} V(\mathbf{x}, \mathbf{y}) \right) \nonumber \\
 &= \dfrac{1}{\gamma} \partial_{\alpha} \left( \rho(\mathbf{x},t) \int d\mathbf{y} \rho(\mathbf{y},t)  \partial_{\alpha} V(\mathbf{x}, \mathbf{y}) \right),
 \label{eq:term_1}
\end{align}
where we have used Eq.~\ref{eq:density_1_part} and $f(\mathbf{x}_{i}) \delta(\mathbf{x} - \mathbf{x}_i) = f(\mathbf{x}) \delta(\mathbf{x} - \mathbf{x}_i)$, which is a property of the delta function. Consider the third term on the RHS of Eq.~\ref{eq:drho_dt_1},
\begin{align}
 \sum_{i} D \, \partial_{\alpha} \partial_{\alpha} \left( \rho_{i}(\mathbf{x},t) \right) &= D \, \partial_{\alpha} \partial_{\alpha} \rho(\mathbf{x},t).
 \label{eq:term_3}
\end{align}
Lastly, we consider the second term on the RHS of Eq.~\ref{eq:drho_dt_1},
\begin{equation}
 \pi(\mathbf{x}, t) \equiv - \sqrt{ 2D} \sum_{i} \partial_{\alpha} \left( \rho_{i}(\mathbf{x},t) \ \eta_{i,\alpha}(t) \right) 
 \label{eq:term_2}
\end{equation}
The correlation of $\pi(\mathbf{x}, t)$ is given as,
\begin{align}
 \avg{\pi(\mathbf{x}, t) \pi(\mathbf{y}, t')} 
 &= 2D \, \sum_i \sum_j \partial_{x, \alpha} \partial_{y, \beta} \, \Bigg[ \rho_{i}(\mathbf{x},t) \rho_{j}(\mathbf{y}, t') \left< \eta_{i,\alpha}(t) \eta_{j,\beta}(t') \right> \Bigg], \nonumber \\
 &= 2D \, \partial_{x, \alpha} \partial_{y, \alpha} \, \Bigg[ \delta (\mathbf{x} - \mathbf{y})  \rho(\mathbf{x},t) \Bigg\{  \delta(t-t') + \dfrac{1}{\sum_{k=0}^{q} \theta_k^2} \sum_{s=1}^{q} \left( \sum_{k=0}^{q-s} \theta_k \theta_{k+s} \right) \left[ \delta(t-t'-s\tau) + \delta(t-t'+s\tau) \right] \Bigg\} \Bigg],
 \label{eq:term_2_cov}
\end{align}
where we have used Eq.~\ref{eq:cov_eta_continuous} and the property of the delta function that $\delta(\mathbf{x} - \mathbf{x}_i) \delta(\mathbf{y} - \mathbf{x}_i) = \delta(\mathbf{x} - \mathbf{x}_i) \delta(\mathbf{x} - \mathbf{y})$. We now define a Gaussian noise field $\chi(\mathbf{x},t)$ as,
\begin{equation}
  \chi(\mathbf{x},t) = \sqrt{2D} \ \partial_{\alpha} \left( \sqrt{\rho(\mathbf{x},t)} \ \psi_{\alpha}(\mathbf{x}, t) \right),
 \label{eq:global_noise_Lambda_div}
\end{equation}
where $\boldsymbol{\psi}(\mathbf{x}, t)$ is a Gaussian noise with mean $\left< \psi_{\alpha}(\mathbf{x}, t) \right> = 0$ and correlation,
\begin{align}
 \left< \psi_{\alpha}(\mathbf{x}, t) \,\psi_{\beta}(\mathbf{y}, t')\right> = \delta_{\alpha \beta} \delta(\mathbf{x} - \mathbf{y}) \left[ \delta(t-t') + \dfrac{1}{\sum_{l=0}^{q} \theta_l^2} \sum_{s=1}^{q}  \left( \sum_{l=0}^{q-s} \theta_l \theta_{l+s} \right) \left[ \delta(t-t'-s\tau) + \delta(t-t'+s\tau) \right] \right].
 \label{eq:psi_correlation_maq}
\end{align}
The correlation function of ${\chi}(\mathbf{x},t)$ is then given by,
\begin{align}
\avg{{\chi}(\mathbf{x}, t) {\chi}(\mathbf{y}, t')}  
 = 2D \, \partial_{x, \alpha} \partial_{y, \alpha} \, \Bigg[ \delta (\mathbf{x} - \mathbf{y})  \rho(\mathbf{x},t) \Bigg\{  \delta(t-t') + \dfrac{1}{\sum_{k=0}^{q} \theta_k^2} \sum_{s=1}^{q} \left( \sum_{k=0}^{q-s} \theta_k \theta_{k+s} \right) \left[ \delta(t-t'-s\tau) + \delta(t-t'+s\tau) \right] \Bigg\} \Bigg].
\label{eq:global_noise_cov}
\end{align}
Eq.~\ref{eq:global_noise_cov} is the same as Eq.~\ref{eq:term_2_cov}. Thus the noise terms ${\pi}(\mathbf{x}, t)$ and ${\chi}(\mathbf{x}, t)$ are identical, allowing us to replace $\pi(\mathbf{x},t)$ with $ {\chi}(\mathbf{x},t)$.
Finally, combining Eqs.~\ref{eq:term_1}, \ref{eq:term_3}, and \ref{eq:global_noise_Lambda_div}, we can write the equation for global density evolution as,
\begin{align}
 \dfrac{\partial \rho(\mathbf{x}, t)}{\partial t} 
 &= \underbrace{ \nabla \cdot \left[ \rho(\mathbf{x}) \left( \dfrac{1}{\gamma} \int \rho(\mathbf{y}) \nabla V(\mathbf{x},\mathbf{y}) d\mathbf{y} \right) \right]}_{\text{drift term}} + \underbrace{ D \nabla^2 \rho(\mathbf{x})}_{\text{diffusion term}} + \underbrace{ \sqrt{2D } \ \nabla \cdot \left( \sqrt{\rho(\mathbf{x})} \ \boldsymbol{\psi}(\mathbf{x}, t) \right)}_{\text{noise term}}.
 \label{eq:drho_dt_2} 
\end{align} 
Eq.~\ref{eq:drho_dt_2} is Eq.~\ref{eq:generic_drho} of the main text.
Notice that plugging $\theta_k = 0 \ \forall \ k \geq 1$ reduces Eq.~\ref{eq:drho_dt_2} to the standard Dean's equation for interacting Brownian particles at equilibrium \cite{dean1996langevin}.

\textbf{Linearization.} We now linearize the density to first order around a spatiotemporally constant mean density $\bar{\rho}$ as $\rho (\mathbf{x}, t) = \bar{\rho} + \delta \rho (\mathbf{x}, t)$.
We substitute this expansion into Eq.~\ref{eq:drho_dt_2}, and retain up to the first-order contributions for the deterministic terms (the first two terms on the RHS of Eq.\ref{eq:drho_dt_2}). For the noise term (the third term on the RHS of Eq.~\ref{eq:drho_dt_2}), we instead only keep the zeroth-order contributions, since all higher-order contributions become $\mathcal{O} (\delta \rho^2)$ when evaluating correlations \cite{dean2016nonequilibrium, dean2014relaxation, demery2014generalized, dinelli2024fluctuating}. It is important to note that while the zeroth-order contribution vanishes for the deterministic terms (the first two terms on the RHS), it remains nonzero for the noise contribution (the third term on the RHS of Eq.~\ref{eq:drho_dt_2}). We can now write the evolution for the density fluctuations $\delta \rho (\mathbf{x}, t)$ as,
\begin{align}
 \dfrac{\partial \delta \rho(\mathbf{x}, t)}{\partial t} 
 &= \dfrac{\bar{\rho}}{\gamma} \ \partial_{\alpha} \left( \int \delta \rho(\mathbf{y}, t) \partial_{\alpha} V(\mathbf{x} - \mathbf{y}) \, d\mathbf{y} \right) + 
 D \partial_{\alpha} \partial_{\alpha} \left( \delta \rho(\mathbf{x}, t) \right) + \sqrt{2D \bar{\rho} } \, \partial_{\alpha} \left( \psi_{\alpha}(\mathbf{x}, t) \right). 
 \label{eq:d_deltarho_dt}
\end{align}
We next define the spatiotemporal Fourier transform for a function $f(\mathbf{x}, t)$ as $\hat{f}(\mathbf{k}, \omega) = \int \int f(\mathbf{x}, t) e^{-i \mathbf{k}\cdot\mathbf{x}} e^{-i \omega t} d \mathbf{x} dt$. Taking the Fourier transform of Eq.~\ref{eq:d_deltarho_dt}, we get,
\begin{align}
 i \omega \hat{\delta \rho} (\mathbf{k}, \omega)
 &= - \dfrac{\bar{\rho}}{\gamma} |\mathbf{k}|^2 \hat{\delta \rho}(\mathbf{k}, \omega) \hat{V}(\mathbf{k}) - 
 D |\mathbf{k}|^2 \hat{\delta \rho}(\mathbf{k}, \omega) + i k_{\alpha} \sqrt{2D \bar{\rho}} \, \hat{\psi}_{\alpha}(\mathbf{k}, \omega),
 \label{eq:deltarho_ft_1}
\end{align}
Rearranging Eq.~\ref{eq:deltarho_ft_1}, we get,
\begin{equation}
 \hat{\delta \rho} (\mathbf{k}, \omega)
 = \dfrac{i k_{\alpha} \sqrt{2D \bar{\rho}} \, \hat{\psi}_{\alpha}(\mathbf{k}, \omega)} {i\omega + |\mathbf{k}|^2 \left[ \dfrac{\bar{\rho} \, \hat{V}(\mathbf{k})}{\gamma} + D \right]}.
 \label{eq:deltarho_ft_2}  
\end{equation}

\textbf{Structure factor.} Let $f^*$ be the complex conjugate of a function $f$. We can then write the dynamic structure factor $S(\mathbf{k}, \omega)$ as,
\begin{align}
 S(\mathbf{k}, \omega) 
 &= \dfrac{1}{2 \pi T N} \left< \hat{\delta \rho} (\mathbf{k}, \omega) \hat{\delta \rho}^* (\mathbf{k}, \omega) \right> =  \dfrac{ D |\mathbf{k}|^2 \left( 1 + 2 \sum_{s=1}^{q} \dfrac{(\sum_{l=0}^{q-s} \theta_l \theta_{l+s})}{\sum_{j=0}^{q} \theta_j^2} \cos(s\omega\tau) \right) } {\pi \left( \omega^2 + \left(|\mathbf{k}|^2 \left[ \dfrac{\bar{\rho} \, \hat{V}(\mathbf{k})}{\gamma} + D \right]\right)^2 \right) }.
\label{eq:sk_dyn} 
\end{align}
Here, $\int dt = T$ denotes the total observation time, $N = \int \rho (\mathbf{x}, t) \, d \mathbf{x}$ is the number of particles, and we have used the relation $\bar{\rho} = N/\int d \mathbf{x}$. We can now write the static structure factor $S(\mathbf{k})$, after some algebra, as,
\begin{align}
 S(\mathbf{k}) 
 &= \int S(\mathbf{k}, \omega) \, d\omega 
    = \dfrac{1}{\left[ \dfrac{\bar{\rho}}{\gamma D} \hat{V}(\mathbf{k}) + 1 \right]} \left( 1 + \dfrac{2}{\sum_{j=0}^q \theta_j^2} \sum_{s=1}^{q} \sum_{l=0}^{q-s} \theta_l \theta_{l+s} \ e^{-s \tau |\mathbf{k}|^2 \left[ \frac{\bar{\rho} \, \hat{V}(\mathbf{k})}{\gamma} + D \right]} \right).
\label{eq:sk_static_1} 
\end{align}
Notice that taking $\theta_k = 0 \ \forall \ k \geq 1$ reduces the system to a system of interacting Brownian particles at equilibrium (see Eqs.~\ref{eq:generic_sde_pgd} and \ref{eq:cov_eta_continuous}), giving,
\begin{align}
 S_{\text{eq}}(\mathbf{k}) 
    &= \dfrac{1}{\left[ \dfrac{\bar{\rho} \, \hat{V}(\mathbf{k})}{\gamma  D} + 1 \right]}.
\label{eq:sk_eqbm} 
\end{align}
Eq.~\ref{eq:sk_eqbm} is the well-known result for interacting Brownian particles at equilibrium, obtained using the random phase approximation \cite{hansen2013theory}. It is evident from Eq.~\ref{eq:sk_static_1} that $S(\mathbf{k})$ depends on the specific functional form of $\hat{V}(\mathbf{k})$. Since we are interested in the long-range ($|\mathbf{k}| \to 0$) behavior of $S(\mathbf{k})$, we now approximate $\hat{V}(\mathbf{k})$ as,
\begin{align}
    \hat{V}(\mathbf{k})
    = \int V(\mathbf{r}) e^{-i\mathbf{k} \cdot \mathbf{r}} d\mathbf{r}
    &\approx \int V(\mathbf{r}) \left( 1 - ik_{\alpha}r_{\alpha} - \dfrac{k_{\alpha}r_{\alpha}k_{\beta}r_{\beta}}{2} + \mathcal{O}(k^3) \right) d\mathbf{r} \nonumber \\
    &= \int V(r) d\mathbf{r} - \dfrac{k_{\alpha}k_{\beta}}{2} \int V(r) r^2 n_{\alpha} n_{\beta} d\mathbf{r} + \mathcal{O}(k^4) \nonumber \\
    &= \underbrace{S_{d} \int V(r) r^{d-1} dr}_{V_1} - k^2 \underbrace{\dfrac{S_{d}}{2d} \int V(r) r^{d+1} dr}_{V_2} + \mathcal{O}(k^4) \nonumber \\
    &= V_1 - k^2 V_2 + \mathcal{O}(k^4),
\label{eq:V_ft}  
\end{align}
where $k = |\mathbf{k}|$, $r = |\mathbf{r}|$, $\mathbf{n} = \mathbf{r}/|\mathbf{r}|$ is the unit vector, $S_d$ is the surface area of a unit $d$-dimensional hypersphere, and $V_1$ and $V_2$ are constants. We have additionally used the fact that $V(\mathbf{r})$ only depends on $r$ and is an even function of $r$ (see Eq.~\ref{eq:v_ji_family_SI}). Plugging Eq.~\ref{eq:V_ft} into Eq.~\ref{eq:sk_static_1}, we get, after some algebra, 
\begin{align}
S(k) = \dfrac{1}{\left(\dfrac{\bar{\rho} V_1}{D \gamma} +  1\right) \sum_{j=0}^q \theta_j^2 } \left[ \left(\sum_{j=0}^q \theta_j\right)^2
+ \left[ \dfrac{ \bar{\rho} V_2 \left(\sum_{j=0}^q \theta_j\right)^2}{\gamma D \left(\dfrac{\bar{\rho} V_1}{D \gamma} + 1 \right)} - 2 D \tau \left(\dfrac{\bar{\rho} V_1}{D \gamma} +  1\right) \sum_{s=1}^{q} s \sum_{l=0}^{q-s} \theta_l \theta_{l+s} \right] k^2 \right] + \mathcal{O}({k}^4).
\label{eq:sk_static_expanded_maq}
\end{align}
Using the definition of Pearson correlation coefficient between the total noise $\zeta_{i, \alpha}^m$ and $\zeta_{i, \alpha}^{m-h}$ (Eq.~\ref{eq:corr_coeff}), we can simplify $S(k)$ to get,
\begin{align}
S(k) = \dfrac{1}{\left(\dfrac{\bar{\rho} V_1}{D \gamma} +  1\right) } \left[ \left( 1 + 2 \sum_{s=1}^q c_s \right)
+ \left[ \dfrac{ \bar{\rho} V_2 \left( 1 + 2 \sum_{s=1}^q c_s \right)}{\gamma D \left(\dfrac{\bar{\rho} V_1}{D \gamma} + 1 \right)} - 2 D \tau \left(\dfrac{\bar{\rho} V_1}{D \gamma} +  1\right) \sum_{s=1}^{q} s c_s \right] k^2 \right] + \mathcal{O}({k}^4).
\label{eq:sk_simplified_full}
\end{align}
Notice that the sum $\sum_{s=1}^q c_s \in [-1/2, q/2]$ (Eq.~\ref{eq:sum_c_bounds}). 
As evident from Eq.~\ref{eq:sk_simplified_full}, when $\sum_{m=1}^q c_m = -1/2$, the structure factor $S(k) \sim k^2$, and the system becomes strongly hyperuniform (class I \cite{torquato2018hyperuniform}). Finally, for the specific case of MA($1$) noise ($q=1$), Eq.~\ref{eq:sk_simplified_full} reduces to, 
\begin{align}
S(k) = \dfrac{1}{\left(\dfrac{\bar{\rho} V_1}{D \gamma} +  1\right) } \left[ \left( 1 + 2 c \right)
+ \left[ \dfrac{ \bar{\rho} V_2 \left( 1 + 2 c \right)}{\gamma D \left(\dfrac{\bar{\rho} V_1}{D \gamma} + 1 \right)} - 2 D \tau \left(\dfrac{\bar{\rho} V_1}{D \gamma} +  1\right) c \right] k^2 \right] + \mathcal{O}({k}^4),
\label{eq:sk_simplified_MA1}
\end{align}
where $c \equiv c_1$. We normalize $S(k)$ using its value at $k \to 0$ for the case of uncorrelated noise ($c=0$) to get $\tilde{S}(k) = S(k)/S_0$, where 
\begin{equation}
    S_0 = \dfrac{1}{\left(\dfrac{\bar{\rho} V_1}{D \gamma} +  1\right) }.
\label{eq:s0}  
\end{equation}
We also normalize $k$ using the value of $k$ for which $\tilde{S} (c=-1) = \lim_{k\to0} \tilde{S} (c=0) = 1.0$ to get $\tilde{k} = k/k_0$, where 
\begin{equation}
    k_0 = \sqrt{ \dfrac{S_0}{D \tau}}.
\label{eq:k0}  
\end{equation}
We can now write Eq.~\ref{eq:sk_simplified_MA1} as,
\begin{align}
    \tilde{S}(\tilde{k}) 
    = (1+2c) + \left[ \underbrace{\dfrac{\bar{\rho} V_2}{\gamma \tau D^2 \left(\dfrac{\bar{\rho} V_1}{D \gamma} +  1\right)^2}}_{B} \left( 1+2c \right) - 2c \right] \tilde{k}^2 + \mathcal{O}(\tilde{k}^4) = (1+2c) + [B(1+2c) - 2c] \tilde{k}^2 + \mathcal{O}(\tilde{k}^4),
\label{eq:sk_normalized_MA1}
\end{align}
where $B$ is a system-dependent constant. Eq.~\ref{eq:sk_normalized_MA1} is Eq.~\ref{eq:struc_fact} of the main text. For our system, the constants are given as, 
\begin{align}
    \gamma &= \dfrac{\tau}{\alpha}, \nonumber \\
    D &= \dfrac{\sigma^2}{2\tau}, \nonumber \\
    V_1 &= \dfrac{\mathcal{E} S_d (2R)^d \Gamma(p+1) \Gamma(d)}{p\Gamma(p+d+1)}, \nonumber \\
    V_2 &= \dfrac{\mathcal{E} S_d (2R)^{d+2} \Gamma(p+1) \Gamma(d+2)}{2 d p \ \Gamma(p+d+3)},
    \label{eq:consts_sgd}  
\end{align}
where $\Gamma$ denotes the Gamma function. Using the definition of $B$, we get,
\begin{equation}
    B = \dfrac{ \dfrac{4\alpha\,\bar{\rho}}{\sigma^{4}}
    \left(\dfrac{\mathcal{E}\,S_d\,(2R)^{d+2}\,\Gamma(p+1)\,\Gamma(d+2)}{2d p\,\Gamma(p+d+3)}\right)}
    {\left(1+ \dfrac{2\alpha\,\bar{\rho} \, \mathcal{E}\,S_d\,(2R)^{d}\,\Gamma(p+1)\,\Gamma(d)}{\sigma^{2}\,p\,\Gamma(p+d+1)}\right)^{2}}.
\label{eq:B_pgd}  
\end{equation}

\section{Stochastic gradient descent (SGD)}

SGD is a well-known optimization algorithm used to train neural networks \cite{bishop2023deep}. In the context of particle systems, SGD corresponds to randomly selecting a subset of terms in the total energy $E = \sum_i \sum_{j \geq i} V_{ij}$, and updating only the corresponding particle positions---either one or both at a time---to minimize this partial energy \cite{zhang2024absorbing, anand2025emergent}. The position $\mathbf{x}_{i}^{m}$ of particle $i$ at time-step $m$ evolves as \cite{zhang2024absorbing, anand2025emergent},
\begin{equation}
    \mathbf{x}_{i}^{m} = \mathbf{x}_{i}^{m-1} - \alpha \sum_{j \in \Gamma_i^{m-1}} \theta_{ji}^{m-1} \nabla_{\mathbf{x}_{i}} V_{ji}^{m-1}.
\label{eq:discrete_sgd_SI}
\end{equation}
Here, $\alpha$ is the learning rate, $V_{ji}^{m-1} = V(|\mathbf{x}^{m-1}_{j} - \mathbf{x}^{m-1}_{i}|)$ is the interaction potential, and $\theta_{ji}^{m-1} \in \{0, 1\}$ is a random variable drawn from a Bernoulli distribution with parameter $b_f$ (batch fraction) at time-step $m-1$. As evident from Eq.~\ref{eq:discrete_sgd_SI}, particles with no overlapping neighbors have no dynamics, and those with at least one overlapping neighbor are called ``active''. $\theta_{ji}^m$ introduces the selection noise in the dynamics and controls the average fraction of active particle pairs $(i,j)$ that move at any given time-step. Further, the selection noise is also pairwise correlated, i.e., $\theta_{ij}^m$ and $\theta_{ji}^m$ are correlated.
Then, the covariance of the full noise vector $\boldsymbol{\omega}^m_{ij} = - \alpha \theta_{ij}^m \nabla_{i} V_{ij}^{m}$ is $\text{cov}[ \omega_{ij, \alpha}^m \omega_{kl, \beta}^n] = \alpha^2 b_f(1-b_f) \, \delta^{mn} \partial_{i,\alpha} V_{ji}^{m} \, \partial_{i,\beta} V_{ji}^{n} (\delta_{ik}\delta_{jl} + c^s \, \delta_{il}\delta_{jk})$, where $c^s$ denotes the Pearson correlation coefficient between $\omega_{ij, \alpha}^m$ and $\omega_{ji, \alpha}^m$. $c^s=0$ implies that $\theta_{ij}^m$ and $\theta_{ji}^m$ are uncorrelated, while $c^s=-1$ corresponds to $\theta_{ij}^m = \theta_{ji}^m$, since $\nabla_{i} V_{ji}^{m} = -\nabla_{j} V_{ji}^{m}$. Notice that the $c^s=-1$ dynamics conserve the pairwise center of mass. 

The discrete-time dynamics of SGD, containing the selection noise, can be approximated by a continuous-time SDE having spatial noise \cite{zhang2024absorbing, anand2025emergent}. It was recently shown that that the long-range structure of SGD, quantified by the structure factor $S(k)$, is given by \cite{anand2025emergent},
\begin{equation}
    \tilde{S}(\tilde{k}) = (1+c^s) + [M(1+c^s) - c^s] \tilde{k}^2,
    \label{eq:sk_sgd}
\end{equation}
where $M$ is a system-dependent constant \cite{anand2025emergent}. For $c^s = -1$, $S(k) \sim k^2$, and the system becomes strongly hyperuniform. 

\section{Stochastic perturbed gradient descent (SPGD)}

Perturbed gradient descent (PGD) dynamics contains temporal noise, given by Eq.~\ref{eq:discrete_pgd_SI}. In contrast, SGD dynamics has selection noise (Eq.~\ref{eq:discrete_sgd_SI}) \cite{zhang2024absorbing, anand2025emergent}. We combine these two distinct noise sources, leading to SPGD dynamics, given by,
\begin{equation}
  \mathbf{x}_{i}^{m} = \mathbf{x}_{i}^{m-1} - \underbrace{ \alpha \sum_{j \in \Gamma_i^{m-1}} \theta_{ji}^{m-1} \, \nabla_{\mathbf{x}_{i}} V_{ji}^{m-1} }_{\text{selection noise term}} + \underbrace{ \dfrac{\sigma}{\Theta} \sum_{k=0}^{q} \theta_k \boldsymbol{\xi}_i^{m-k} }_{\text{temporal noise term}},
\label{eq:discrete_spgd_SI}
\end{equation}
where all symbols are defined as in Eqs.~\ref{eq:discrete_pgd_SI} and \ref{eq:discrete_sgd_SI}. SPGD dynamics typically lead to ``good'', i.e. generalizable solutions in neural networks compared to either PGD or SGD alone \cite{orvieto2022anticorrelated}. We now approximate the discrete-time dynamics of SPGD by a continuous-time SDE. Following the steps outlined in Sec.~I and in ref.~\cite{anand2025emergent}, we get,
\begin{equation}
 \dfrac{d\mathbf{x}_{i}(t)}{dt}
 = \underbrace{ - \dfrac{1}{\gamma} \sum_{j = 1}^N \nabla_{\mathbf{x}_i} V_{ji}}_{\text{deterministic term}} 
 + \underbrace{ \sum_{j = 1}^N \sqrt{\mathbf{\Lambda}_{ji}} \cdot \boldsymbol{\kappa}_{ji}}_{\text{spatial noise term}} + \underbrace{ \sqrt{2D} \ \boldsymbol{\eta}_{i} (t)}_{\text{temporal noise term}},
 \label{eq:generic_sde_spgd} 
\end{equation}
where $\gamma = \tau/\alpha b_f$, $D = \sigma^2/2 \tau$, $\Lambda_{ji, \alpha \beta} = (\alpha^2 b_f (1 - b_f)/\tau) \, \partial_{\alpha} V_{ji} \partial_{\beta} V_{ji}$ is the projection matrix which aligns the noise along the line joining the center of a pair of overlapping particles, and $\sqrt{.}$ denotes the matrix square root in the second term on the RHS. $\boldsymbol{\kappa}_{ji}$ is a pairwise Gaussian noise given by the particle $j$ to particle $i$ with mean $\left< {\kappa}_{ji,\alpha} (t) \right> = 0$ and the covariance matrix $\left< {\kappa}_{ij,\alpha} (t) \, {\kappa}_{kl,\beta} (t') \right> = \delta (t-t') \, \delta_{\alpha \beta} (\delta_{ik}\delta_{jl} + c^s \, \delta_{il}\delta_{jk})$, where $c^s \in [-1,0]$ denotes the Pearson correlation coefficient between ${\kappa}_{ij,\alpha} (t)$ and ${\kappa}_{ji,\alpha} (t)$. The spatial noise $\boldsymbol{\kappa}_{ji}$ and the temporal noise $\boldsymbol{\eta}_{i}$ are uncorrelated, i.e., $\left< {\kappa}_{ij,\alpha}(t) \,{\eta}_{k,\beta}(t') \right> = 0_{\alpha \beta}$.

We now aim to coarse-grain the continuous-time approximation of SPGD. Following the method described in Sec.~II and ref.~\cite{anand2025emergent}, we get the equation for density evolution as,
\begin{align}
    \dfrac{\partial \rho(\mathbf{x}, t)}{\partial t} 
    &= \underbrace{ \nabla \cdot \left[ \rho(\mathbf{x}) \left( \dfrac{1}{\gamma} \int \rho(\mathbf{y}) \nabla V(\mathbf{x},\mathbf{y}) d\mathbf{y} \right) \right]}_{\text{drift term}} + \underbrace{ \nabla \nabla : \left( \rho(\mathbf{x}) \int \rho(\mathbf{y}) {\mathbf{\Lambda}(\mathbf{x}, \mathbf{y})} d\mathbf{y}  \right) }_\text{diffusion term $1$} + \underbrace{ D \nabla^2 \rho(\mathbf{x})}_{\text{diffusion term $2$}} \nonumber \\
    &\quad+ \underbrace{\nabla \cdot \left[ \sqrt{\rho(\mathbf{x})} \int \sqrt{\rho(\mathbf{y})} \sqrt{\boldsymbol{\Lambda}(\mathbf{x}, \mathbf{y})} \cdot \boldsymbol{\epsilon}(\mathbf{x}, \mathbf{y}, t) d \mathbf{y} \right] }_\text{noise term $1$} + \underbrace{ \sqrt{2D } \ \nabla \cdot \left( \sqrt{\rho(\mathbf{x})} \ \boldsymbol{\psi}(\mathbf{x}, t) \right)}_{\text{noise term $2$}},    
    \label{eq:drho_dt_spgd}
\end{align} 
where $:$ denotes the double dot product. $\boldsymbol{\epsilon}(\mathbf{x}, \mathbf{y}, t)$ is a two-point, Gaussian noise field having mean $\left< {\epsilon}_{\alpha}(\mathbf{x}, \mathbf{y}, t) \right> = 0$, and the covariance matrix $\left< {\epsilon}_{\alpha}(\mathbf{x}, \mathbf{y}, t) \, {\epsilon}_{\beta}(\mathbf{u}, \mathbf{w}, t')\right> = \delta_{\alpha \beta} \delta(t-t') \, [ \delta(\mathbf{x} - \mathbf{u}) \, \delta(\mathbf{y} - \mathbf{w}) + \, c^s \, \delta(\mathbf{x} - \mathbf{w}) \, \delta(\mathbf{y} - \mathbf{u}) ]$ \cite{anand2025emergent}. $c^s$ is the Pearson correlation coefficient between $\epsilon_{\alpha}(\mathbf{x}, \mathbf{y}, t)$ and $\epsilon_{\alpha}(\mathbf{y}, \mathbf{x}, t)$. Following the linearization described in Sec.~II and in ref.~\cite{anand2025emergent}, we can write $S(k)$, after some algebra, as,
\begin{align}
    S(\mathbf{k}) = \dfrac{ 
    \dfrac{\overline{\rho}}{2}\, k_\alpha k_\beta \left( A_{\alpha\beta} + c\,\hat{\Lambda}_{\alpha\beta} (\mathbf{k}) \right)
    + D k^2
    + 2D k^2 \sum_{s=1}^{q} c_s \, \exp\left\{-s \tau \left( 
    \frac{\overline{\rho}\,\hat{V} (\mathbf{k})}{\gamma}\,k^2 
    + \frac{\overline{\rho}}{2}\, k_\alpha k_\beta \left( \hat{\Lambda}_{\alpha\beta} (\mathbf{k}) + A_{\alpha\beta} \right)
    + D k^2 
    \right) \right\}
    }{
    \dfrac{\overline{\rho}\,\hat{V} (\mathbf{k})}{\gamma}\,k^2 
    + \dfrac{\overline{\rho}}{2}\, k_\alpha k_\beta \left( \hat{\Lambda}_{\alpha\beta} (\mathbf{k}) + A_{\alpha\beta} \right)
    + D k^2 
    },
    \label{eq:sk_spgd}
\end{align}
where $A_{\alpha \beta} = \int {\Lambda(\mathbf{r})}_{\alpha \beta} \, d\mathbf{r} = A_1 \delta_{\alpha \beta}$ \cite{anand2025emergent}. Following the long-wavelength expansion of $\hat{\Lambda}_{\alpha\beta} (\mathbf{k})$ (as described in ref.~\cite{anand2025emergent}), and $\hat{V}(\mathbf{k})$ (Eq.~\ref{eq:V_ft}), we get, after some simplification,
\begin{align}
    S(k) &= \frac{1}{M_0} 
    \left[ \frac{\displaystyle \overline{\rho} A_1}{2}(1+c^s) + D \left( 1+2\sum_{m=1}^q c_m \right) \right] \nonumber \\
    &\quad+ \displaystyle \frac{1}{M_0} \left[ \left(\dfrac{\overline{\rho}A_1}{2}(1+c^s) + D \left(1+2 \sum_{m=1}^q c_m \right) \right) \dfrac{M_1}{M_0}  -\dfrac{\overline{\rho} (\lambda_1+3\lambda_2)}{2} c^s - 2D\tau M_0 \sum_{m=1}^q m c_m \right] k^2
    + \mathcal{O}(k^4),
    \label{eq:sk_spgd_simplified}
\end{align}
where $M_0$ and $M_2$ are given by, 
\begin{align}
M_0 &= \frac{\overline{\rho}V_1}{\gamma} + \overline{\rho}A_1 + D \nonumber \\
M_1 &= \frac{\overline{\rho}V_2}{\gamma} + \dfrac{\overline{\rho} (\lambda_1+3\lambda_2)}{2}.
    \label{spgd_constants}
\end{align}
It is evident from Eq.~\ref{eq:sk_spgd_simplified} that when both the spatial and temporal noise are strongly anti-correlated ($c^s = -1$ and $\sum_{m=1}^{q} c_m = -1/2$), the the structure factor $S(k) \sim k^2$, and the system becomes strongly hyperuniform.  For the specific case of MA($1$) noise ($q=1$), Eq.~\ref{eq:sk_spgd_simplified} reduces to, 
\begin{align}
S(k) &= \frac{1}{M_0} 
    \left[ \frac{\displaystyle \overline{\rho} A_1}{2}(1+c^s) + D \left( 1+2c \right) \right] \nonumber \\
    &\quad+ \displaystyle \frac{1}{M_0} \left[ \left(\dfrac{\overline{\rho}A_1}{2}(1+c^s) + D \left(1+2  c \right) \right) \dfrac{M_1}{M_0}  -\dfrac{\overline{\rho} (\lambda_1+3\lambda_2)}{2} c^s - 2D\tau M_0 c \right] k^2
    + \mathcal{O}(k^4),
\label{eq:sk_spgd_simplified_MA1}
\end{align}
where $c \equiv c_1$. Finally, we can normalize $S$ by $S_0$ (Eq.~\ref{eq:s0}) and $k$ by $k_0$ (Eq.~\ref{eq:k0}) to get,
\begin{align}
\tilde{S}(\tilde{k}) &= \frac{\left(\dfrac{\bar{\rho} V_1}{D \gamma} +  1\right)}{M_0} 
    \left[ \frac{\displaystyle \overline{\rho} A_1}{2}(1+c^s) + D \left( 1+2c \right) \right] \nonumber \\
    &\quad+ \displaystyle \frac{1}{D \tau M_0} \left[ \left(\dfrac{\overline{\rho}A_1}{2}(1+c^s) + D \left(1+2  c \right) \right) \dfrac{M_1}{M_0}  -\dfrac{\overline{\rho} (\lambda_1+3\lambda_2)}{2} c^s - 2D\tau M_0 c \right] \tilde{k}^2
    + \mathcal{O}(\tilde{k}^4).
\label{eq:sk_spgd_simplified_MA1_norm}
\end{align}
For SPGD, the constants are given as \cite{anand2025emergent},
\begin{align}
    \gamma^{\text{SPGD}} &= \frac{\tau}{\alpha b_f}, \nonumber \\
    D^{\text{SPGD}} &= \dfrac{\sigma^2}{2\tau}, \nonumber \\
    V_1^{\text{SPGD}} &= \frac{\mathcal{E} S_d (2R)^d \Gamma(p+1) \Gamma(d)}{p\Gamma(p+d+1)}, \nonumber \\
    V_2^{\text{SPGD}} &= \frac{\mathcal{E} S_d (2R)^{d+2} \Gamma(p+1) \Gamma(d+2)}{2 d p \ \Gamma(p+d+3)}, \nonumber \\
    A_1^{\text{SPGD}} &= \frac{S_d \mathcal{E}^2 \alpha^2 b_f (1-b_f) (2R)^d \Gamma(2p-1) \Gamma(d)}{4 R^2 \tau d \ \Gamma(2p+d-1)}, \nonumber
 \\
     \lambda_1^{\text{SPGD}} &= 0, \nonumber \\
     \lambda_2^{\text{SPGD}} &= \frac{S_d \mathcal{E}^2 \alpha^2 b_f (1-b_f) (2R)^{d+2} \Gamma(2p-1) \Gamma(d+2)}{8 d (d+2) R^2 \tau \ \Gamma(2p+d+1)}.
    \label{eq:consts_spgd}  
\end{align}

\bibliographystyle{apsrev4-2}
\bibliography{ref.bib}

\clearpage

\begin{figure}[htbp!]
    \centering
    \includegraphics[width=\linewidth]{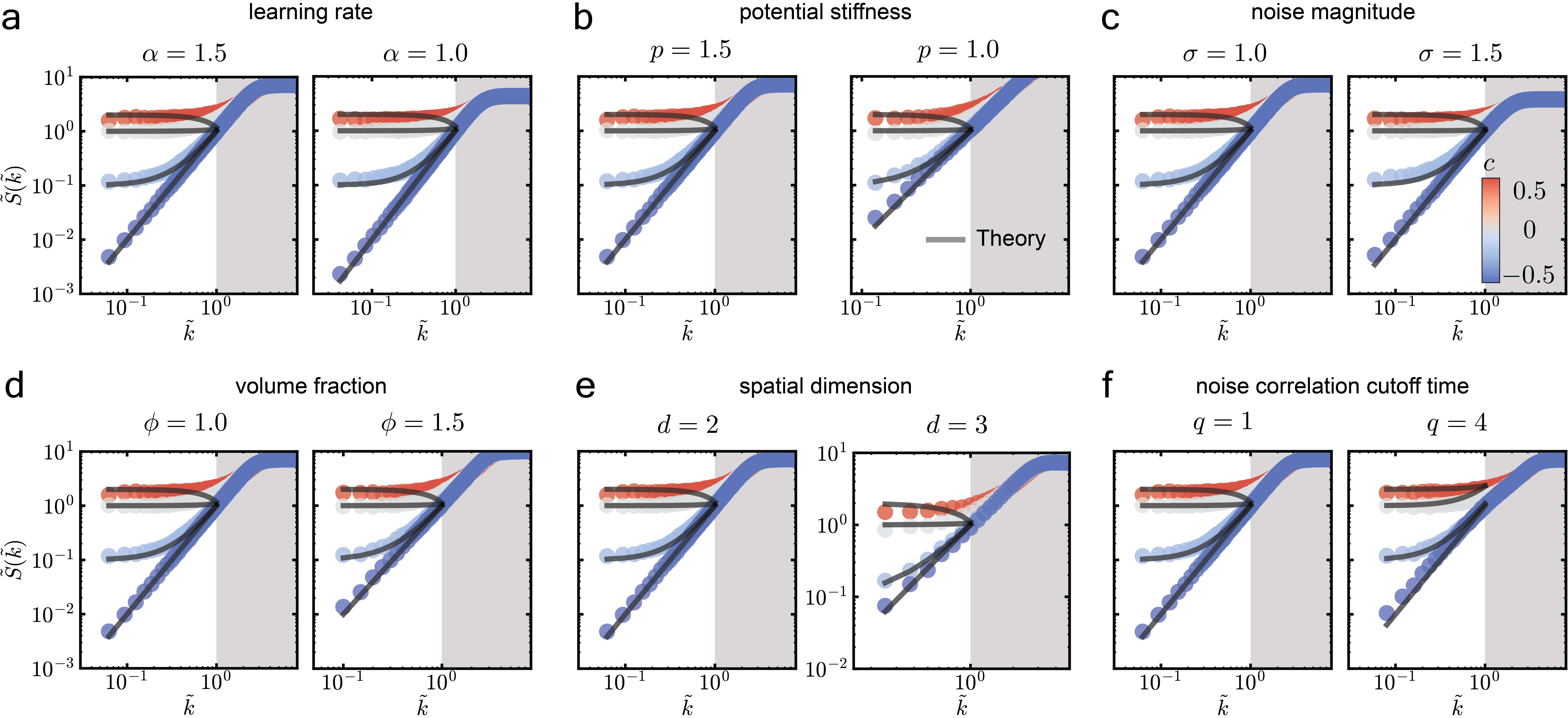}
    \caption{\small \textbf{Dependence of the structure factor on system parameters.} The normalized radially averaged structure factor $\Tilde{S}(\Tilde{k})$ versus normalized radial wave number $\Tilde{k}$ for variations in learning rate $\alpha$ (a), potential stiffness $p$ (b), noise magnitude $\sigma$ (c), particle volume fraction $\phi$ (d), spatial dimension $d$ (e), and the noise correlation cutoff time $q$ (f). Circles denote particle simulations. Baseline parameters are set as $N = 10^5$, $\alpha = 1.5$, $p = 1.5$, $\sigma = 1.0$, $\phi = 1.0$, $d = 2$, $q=1$, and the noise correlation coefficient $c \in [-0.5, 0.5]$. Each panel varies only one parameter. For $q > 1$, $c \equiv \sum_{m=1}^q c_m$. For $d=3$, we choose a bigger system with $N = 10^6$ to better probe large length scales. The normalized structure factor is defined as $\tilde{S} = S(k)/S_{0}(2\pi/L)$ where $S_{0}(2\pi/L)$ is the structure factor at $k = 2\pi/L$ for $c=0$, and $L$ is the side length of the simulation box. The normalized wave number is $\tilde{k} = k/k_0$ where $k_0$ is chosen such that $\tilde{S}(k_0) = 1$ for maximally anti-correlated noise ($c=-0.5$). Black lines in panels (a)-(e) show predictions of Eq.~\ref{eq:sk_normalized_MA1} for different values of $c$, with $B$ substituted from Eq.~\ref{eq:B_pgd}. Black lines in (f) show predictions of the normalized version of Eq.~\ref{eq:sk_simplified_full} for different values of $c \equiv \sum_{m=1}^q c_m$, where we normalize $S(k)$ by $S_0$, and $k$ by $\sqrt{S_0/(-2D\tau \sum_{m=1}^q m c_m)}$ (see Eqs.~\ref{eq:sk_simplified_full} and \ref{eq:s0}). Gray shaded regions correspond to short-range behavior ($\tilde{k}>1$).
}
    \label{fig:figure1_si}
\end{figure}

\clearpage

\begin{figure}[htbp!]
    \centering
    \includegraphics[width=\linewidth]{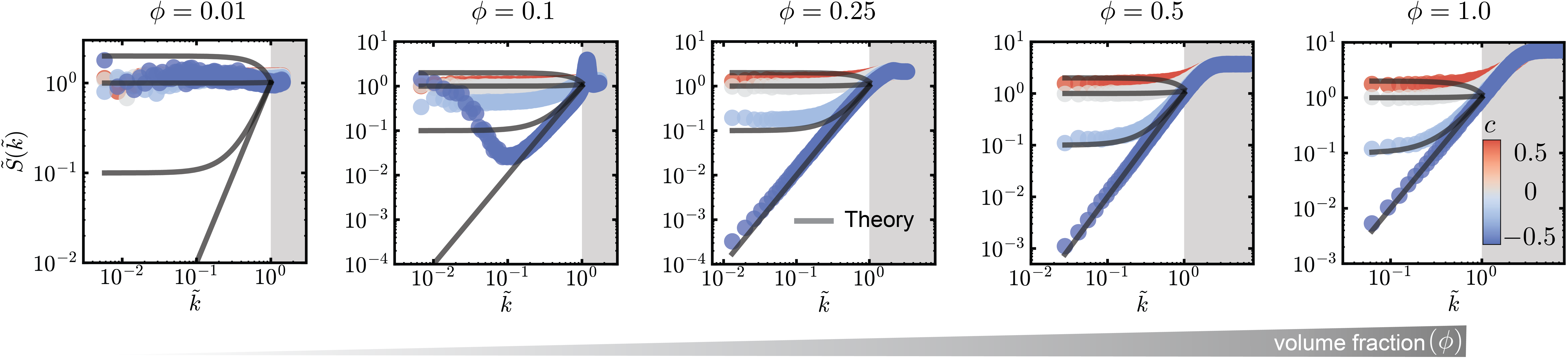}
    \caption{\small \textbf{Dependence of the long-range structure on the volume fraction $\phi$.} The normalized radially averaged structure factor $\Tilde{S}(\Tilde{k})$ versus normalized radial wave number $\Tilde{k}$ for increasing particle volume fraction $\phi$ (left to right). Circles denote particle simulations. The normalized structure factor is defined as $\tilde{S} = S(k)/S_{0}(2\pi/L)$ where $S_{0}(2\pi/L)$ is the structure factor at $k = 2\pi/L$ for $c=0$, and $L$ is the side length of the simulation box. The normalized wave number is $\tilde{k} = k/k_0$ where $k_0$ is chosen such that $\tilde{S}(k_0) = 1$ for maximally anti-correlated noise ($c=-0.5$). All parameters are identical as in the main text (see Methods), except $\phi$, which is varied from $0.01$ to $1.0$. Black lines show predictions of Eq.~\ref{eq:sk_normalized_MA1} for different values of $c$, with $B$ substituted from Eq.~\ref{eq:B_pgd}. Gray shaded regions correspond to short-range behavior ($\tilde{k}>1$).
}
    \label{fig:figure2_si}
\end{figure}

\clearpage

\begin{figure}[htbp!]
    \centering
    \includegraphics[width=\linewidth]{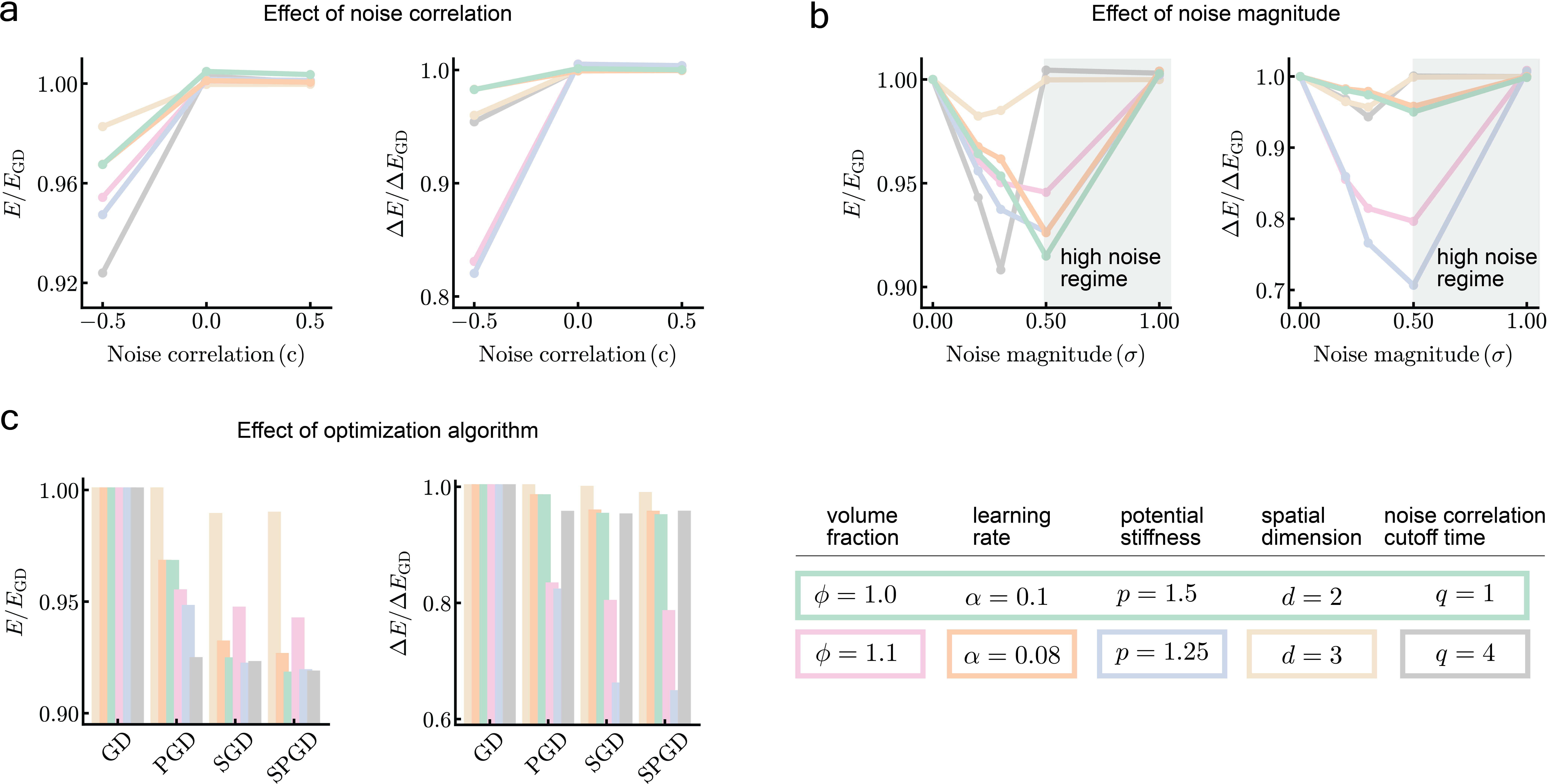}
    \caption{\small \textbf{Dependence of the energy $E$ and the energy change $\Delta E$ upon small perturbation on system parameters.} Starting from a random initial condition, the system undergoes noisy dynamics according to perturbed gradient descent (PGD) (Eq.~\ref{eq:discrete_pgd_SI}), until it reaches a steady-state. The system is then quenched via (noiseless) gradient descent (GD), until it reaches an energy minimum. The system has an energy $E$ at the minimum and an energy $E + \Delta E$ after applying a small perturbation to the system. All parameters are identical as in the main text (see Methods). (a) Normalized $E$ and $\Delta E$ versus noise correlation $c$ in particle simulations. Both quantities are normalized by the corresponding values from (noiseless) GD. For $q > 1$, $c \equiv \sum_{m=1}^q c_m$. (b) Normalized $E$ and $\Delta E$ versus noise magnitude $\sigma$ in particle simulations. Both quantities are normalized by the corresponding values from (noiseless) GD ($\sigma = 0$). For low noise magnitudes, both $E$ and $\Delta E$ decrease as $\sigma$ increases. As $\sigma$ increases further, at sufficiently high noise, all parameter choices ultimately converge to GD. (c) Normalized $E$ and $\Delta E$ for different dynamics in particle simulations, where both quantities are normalized by the corresponding values from GD. We compare (noiseless) GD with three different noisy dynamics: PGD (Eq.~\ref{eq:discrete_pgd_SI}), stochastic gradient descent (SGD) (Eq.~\ref{eq:discrete_sgd_SI}), and stochastic perturbed gradient descent (SPGD) (Eq.~\ref{eq:discrete_spgd_SI}). SPGD combines temporal noise from PGD with selection (or spatial) noise from SGD. For $d=3$, we choose a lower noise magnitude ($\sigma = 0.1$) for PGD and SPGD to stay in the low noise regime (see (b)).
}
    \label{fig:figure3_si}
\end{figure}

\clearpage

\begin{figure}[htbp!]
    \centering
    \includegraphics[width=\linewidth]{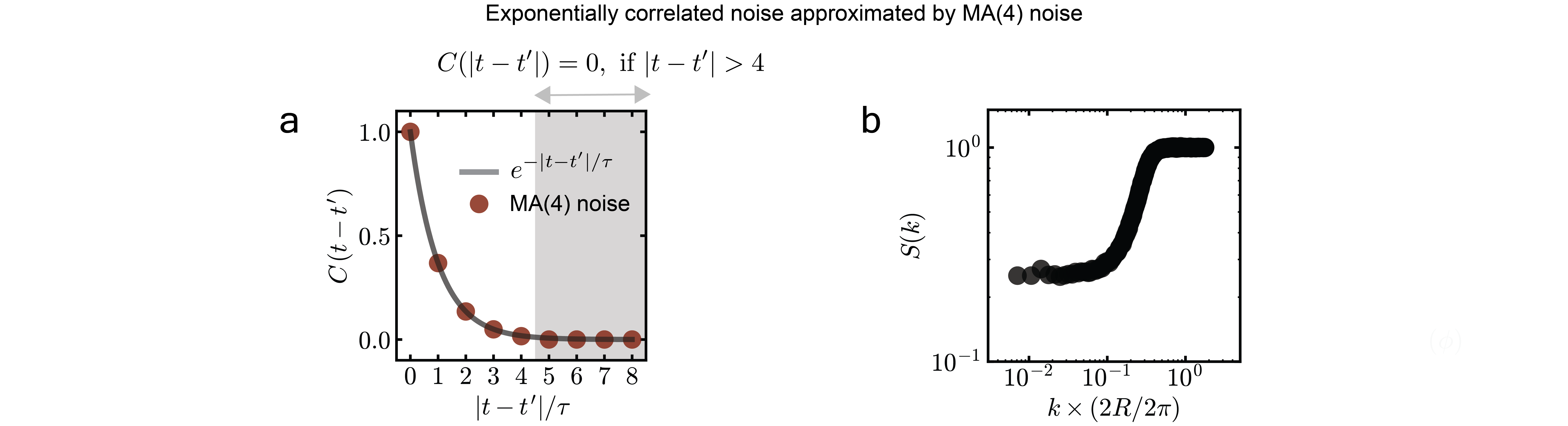}
    \caption{\small \textbf{Long-range structure for exponentially correlated temporal noise.} We use the MA($4$) noise to approximate the exponentially correlated noise. For the MA($4$) noise, the noise correlation $C(|t-t'| = 0$, when $|t-t'| > 4$ (Eq.~\ref{eq:cov_dis}). Here, $C(|t-t'| \equiv c_h$, and $|t-t'| = h \tau$, where $\tau$ is the characteristic time elapsed per step (Eq.~\ref{eq:corr_coeff}). (a) Comparison of noise correlations for the exponentially correlated and the MA($4$) noise. Black line denotes exact exponentially decaying correlations $e^{-|t-t'|/\tau}$. Circles denote correlations of the MA($4$) noise with parameters $\theta_0 = 1, \theta_1 = 0.368, \theta_2 = 0.135, \theta_3 = 0.050, \theta_4 = 0.018$, and $ \theta_k = 0 \ \forall \ k \geq 5$. Gray shaded region corresponds to $C(|t-t'| = 0$ regime. It is evident that the MA($4$) noise can approximate the exponentially decaying correlations. (b) The radially averaged structure factor $S(k)$ versus dimensionless radial wave number $k$ in particle simulations for the MA($4$) noise in (a). $R$ is the particle radius. All parameters are the same as in the main text (see Methods).
}
    \label{fig:figure4_si}
\end{figure}

\clearpage

\begin{figure}[htbp!]
    \centering
    \includegraphics[width=\linewidth]{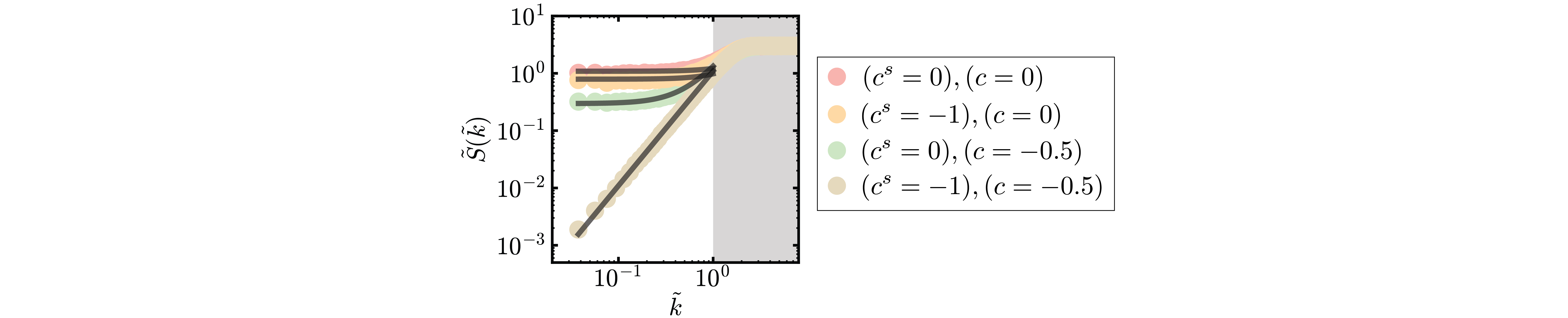}
    \caption{\small \textbf{Long-range structure for stochastic perturbed gradient descent (SPGD).} The normalized radially averaged structure factor $\Tilde{S}(\Tilde{k})$ versus normalized radial wave number $\Tilde{k}$ for SPGD (Eq.~\ref{eq:discrete_spgd_SI}). Circles denote particle simulations. The normalized structure factor is defined as $\tilde{S} = S(k)/S_{0}(2\pi/L)$, where $S_{0}(2\pi/L)$ is the structure factor at $k = 2\pi/L$ for $(c^s = 0, c=0)$, and $L$ is the side length of the simulation box. The normalized wave number is $\tilde{k} = k/k_0$ where $k_0$ is chosen such that $\tilde{S}(k_0) = 1$ for maximally anti-correlated noise ($c^s = -1, c=-0.5$). Black lines show predictions of Eq.~\ref{eq:sk_spgd_simplified_MA1_norm} for different values of the spatial noise correlation coefficient $c^s$ and the temporal noise correlation coefficient $c$. Parameters are set as $N = 10^5$, $\alpha = 1.5$, $p = 1.5$, $\sigma = 1.0$, $\phi = 1.0$, $d = 2$, $q=1$, and $b_f = 0.5$. Gray shaded regions correspond to short-range behavior ($\tilde{k}>1$).}
    \label{fig:figure5_si}
\end{figure}